\documentclass{pasa}%

\usepackage{graphicx}
\pdfoutput=1

\title[Resistive Accretion Flows around a Kerr Black Hole]{Resistive Accretion Flows around a Kerr Black Hole}

\author[M. Shaghaghian]{M. Shaghaghian \thanks{shaghaghian@iaushiraz.ac.ir}
\affil{Department of Physics, Shiraz Branch, Islamic Azad University, Shiraz, Iran}%
}%

\jid{PASA}
\doi{10.1017/pas.\the\year.xxx}
\jyear{\the\year}

\usepackage{aas_macros}
\usepackage{hyperref}
\hypersetup{colorlinks,citecolor=blue,linkcolor=blue,urlcolor=blue}

\hypersetup{draft}

\begin{document}

\begin{frontmatter}
\maketitle

\begin{abstract}
In this paper, we present the stationary axisymmetric configuration of a resistive magnetised
thick accretion disc in the vicinity of external gravity and intrinsic
dipolar magnetic field of a slowly rotating black hole. The plasma is described by the equations of fully general
relativistic magnetohydrodynamics (MHD) along with the Ohm's law and in the absence
of the effects of radiation fields. We try to solve these two-dimensional MHD equations analytically as much as possible. However, we sometimes inevitably refer to numerical methods as well. To fully
understand the relativistic geometrically thick accretion disc
structure, we consider all three components of the fluid velocity to be
non-zero. This implies that the magnetofluid can flow in all three
directions surrounding the central black hole. As we get radially
closer to the hole, the fluid flows faster in all those directions.
However, as we move towards the equator along the meridional
direction, the radial inflow becomes stronger from both the speed
and the mass accretion rate points of view. Nonetheless, the
vertical (meridional) speed and the rotation of the plasma disc
become slower in that direction. Due to the presence of pressure
gradient forces, a sub-Keplerian angular momentum distribution
throughout the thick disc is expected as well. To get a concise analytical form of the rate of accretion, we assume that the radial
dependency of radial and meridional fluid velocities is the same. This simplifying assumption leads to radial independency of mass
accretion rate. The motion of the accreting plasma produces
an azimuthal current whose strength is specified
based on the strength of the external dipolar magnetic field. This
current generates a poloidal magnetic field in the disc which
is continuous across the disc boundary surface due to the presence
of the finite resistivity for the plasma. The gas in the disc is
vertically supported not only by the gas pressure but also by the
magnetic pressure.
\end{abstract}

\end{frontmatter}

\section{INTRODUCTION }
\label{sec:intro}

Transforming gravitational energy into radiation in the most
efficient possible form takes place in accretion onto black holes.
It is certainly believed to be the primary power source behind the
most luminous astrophysical systems that range from quasars and active
galactic nuclei (AGNs) with very massive black holes to X-ray
binaries with stellar-mass black holes. Due to the sufficiently high
angular momentum content of the accreting matter, accretion does not happen
as direct free fall onto a central star. Instead, it is expected to
occur in the form of a disc. For formation of an accretion disc, it is necessary that the angular momentum is extracted from
the inner to the outer regions of the disc. The transport mechanism of
this angular momentum is complicated and not
entirely clear (Lee $\&$ Ruiz 2002). Hence, regardless of the physics behind the cause
responsible for this transport, Shakura $\&$ Sunyaev (1973)
(hereafter SS73) suggested an enormously productive Ansatz about
viscosity that is parameterised with an $\alpha$-parameter. This
parametrisation introduced by the standard model has been quite
successful in interpreting the gross features of observational
results.

It is generally believed that magnetic fields are ubiquitous in
accretion discs and play a relevant role in their physical
scenarios. As a result, it is not unexpected that
the magnetic fields are involved in a broad variety of dynamical
processes of accretion discs. For instance, the magnetic stress may
take the place of viscous stress of the standard model. Similar to
the results of $\alpha$-viscid disc of SS73, in a magnetised disc,
about one-half of the released gravitational energy can dissipate
through the Joule dissipation and the work done against the pressure
force (Kaburaki 1986; 1987).

Knowing that the magnetic stress may drive disc turbulence and the
outward transport of angular momentum, the theory of accretion disc is
moving from the relatively simple parameterised, one-dimensional standard
model of SS73 towards more realistic models. Considering the
electrical conductivity for the plasma as a dissipative factor may
be one of the most effective steps towards this aim. However, the
underlying physics becomes more complex, especially when strong
gravitational and external magnetic fields are also present. On
account of this complexity, some authors intend to employ the limit
of infinite conductivity, that is, the so-called ideal magnetohydrodynamics (MHD).
In a no-resistive plasma, the magnetic lines of force freeze in and
advect with the plasma. Furthermore, conservation of the magnetic
flux passing through a moving surface in a no-resistive magnetofluid
and no change in the topology of the magnetic field lines are of the
other features of an ideal plasma. One important consequence of
these properties is that the crossed magnetic field lines are not permitted to reconnect together in a
perfectly conducting fluid (Eyink $\&$ Aluie 2006).
Ideal MHD approximation applies widely in
many astrophysical relevant situations including both theoretical
models (Banerjee et al. 1997; Yuan $\&$ Narayan 2014) and numerical
simulations (Koide, Shibata, $\&$ Kudoh 1999; De Villiers, Hawley, $\&$ Krolik 2003; McKinney
$\&$ Gammie 2004).

Nevertheless, there are other physical circumstances in which this
approximation no longer holds and it is important to include the
effects of finite conductivity. For instance, in cold, dense plasmas around
protostars (Fleming $\&$ Stone 2003) or dwarf nova systems (Gammie
$\&$ Menou 1998), the ionization fraction is so small.
Furthermore, in the radiatively inefficient accretion flows (Foucart et al. 2016; 2017),
despite the high temperature and fully ionised gas,
the mean free path for coulomb interactions between
charged particles is much larger than the typical size of the
system. The accreting matter, in these cases, is probably expected to be nearly collisionless and
its dynamical evolution may differ significantly from ideal MHD predictions as well. Inclusion of a finite resistivity as
an angular momentum transport mechanism, particularly, is essential
for a non-viscous disc to liberate the gravitational energy (Kaburaki 1986; 1987; Tripathy, Prasanna $\&$ Das 1990;
Banerjee et al. 1995; Kudoh $\&$ Kaburaki 1996; Koide 2010).

Accretion discs surrounding the black holes are among the most
luminous objects in the universe. That is why they are
the centre of special observational attention. They are also of
considerable theoretical interest. To fully understand their
structure and evolution, and to compare accurately the theoretical
results with observational evidences, it is necessary to consider
the influences of plasma interaction with all possible fields. These
fields include a powerful relativistic gravitational field, a strong
electromagnetic field and an intense radiation field (Gammie $\&$
McKinney 2003). In this occasion, one encounters with a coupled set
of partial differential equations that are time-dependent,
multidimensional, and highly non-linear. Those equations may involve a large number of
different physical quantities and free parameters. This, in turn, may be an obstacle on the
way of analytical solutions and even numerical solutions of the
equations due to the limitation of the ability of today's computers.
Consequently, because of the level of complexity that the radiation
field introduces, as the first feasible approximation, it seems
reasonable that one ignores the radiation field and studies the
problem in a non-radiating MHD mode.

The flows in the accretion discs may exhibit different morphologies
from the viewpoint of their geometrical shape. They are generally
divided into two distinct classes, thin discs and thick discs.
Theory of thin accretion discs is well developed and has a fairly
firm observational basis (SS73). However, thick accretion discs
suffer mainly from the lack of an universally accepted model. Besides,
their relevant observations are still rare and indirect. Therefore,
there are still many theoretical uncertainties about their nature
and structure. Nevertheless, observational and theoretical studies
of thick accretion discs are of special astrophysical importance.
Since willy-nilly, such structures have been suggested as models of
quasars, AGNs, some X-ray binaries and are
probably present in the central engine of gamma-ray bursts
(Abramowicz, Karas $\&$ Lanza 1998; Font $\&$ Daigne 2002).

Low-mass X-ray binaries are excellent laboratories for experimenting
accretion physics because they are much closer than the AGNs and
are therefore somewhat easier to observe (Higginbottom et al. 2018).
When the central compact object is a black hole, it is usual that
the dynamics obey the general relativistic. Accordingly, we pursue
the fully general relativistic magnetofluids around a typical black
hole in a low-mass X-ray binary system such as Cygnus X-1. In this
context, we are motivated to put aside the ideal MHD approximation
and investigate the resistive accretion disc structure without
invoking the thin disc approximation in the absence of the effects
of radiation field.

Standard thin disc theory of SS73 is characterised by the
equilibrium of centrifugal and gravitational forces in the radial
direction. It leads to the Keplerian rotation law throughout the
disc. Whereas there is no net gas flow in the vertical direction,
momentum conservation equation in that direction reduces to the
hydrostatic equilibrium equation. Once the thin disc approximation
is laid aside, the vertical thickness becomes comparable with its
radial extension. The pressure gradient forces provide an
essential support in the radial as well as meridional direction. At
this moment, gravity can not be balanced by centrifugal forces any
more. The rotation law is no longer Keplerian and the vertical
hydrostatic equilibrium is abandoned. Another key characteristic of
SS73 theory is its subcritical luminosity. It means that the maximum
possible luminosity of the standard geometrically thin discs is the
Eddington luminosity. That is, the luminosity in which the inward
gravity on accreting fluid is precisely counterbalanced by the
outward radiation pressure gradient force of photons. Evidently, if
the disc's luminosity exceeds the Eddington value, then some matter
will be blown off by the pressure of the supercritical radiation
flux in the form of a wind or a collimated bipolar jet (Okuda 2002;
Takeuchi, Ohsuga $\&$ Mineshige 2010; Takahashi $\&$ Ohsuga 2015).

As a result, we are interested in relativistic accretion tori around
a slowly rotating black hole in the sub-Eddington regime and
non-radiating mode, considering all three components of flow
velocity to be non-zero. Particularly in what concerns the
relativistic geometrically thick accretion discs, most of the
previous works are mainly devoted to equilibrium toroidal
configurations. That is, the flow restricted to the azimuthal
component only, with assumption that the radial and meridional
components are negligible in comparison with the azimuthal one
(Banerjee et al. 1997; Kovar et al. 2011; Trova et al. 2018).
Similar to our idea, both in Newtonian regime (Tripathy, Prasanna
$\&$ Das 1990) and in relativistic regime for a non-rotating Schwarzschild black hole without
the vertical component for the magnetofluid velocity (Shaghaghian 2016), already
have been done.
It is known
that almost all of the celestial bodies have a non-zero spin, and thus,
the Schwarzschild geometry does not tell the whole story. Thus, we
extend this idea to the case of slowly rotating black hole and let
the magnetofluid flows in all three directions.

The structure of this paper is organised as detailed below: We begin
in the next section, with a presentation of the theoretical
framework used to construct our desired model and to describe the
background geometry and the external magnetic field. Also we depict
our disc scheme in this section. The general formalism of the
problem is discussed in Section 3. It includes the basic equations
governing the relativistic magnetised flow accreted from the plasma
around a slowly rotating black hole in the form of a thick torus, as
well as their self-consistent solutions along with the physical
simplifications of the problem. Our main conclusions are summarised
in Section 4.

\section{The model}
\subsection{Spacetime}
To investigate the relativistic accretion flows around a rotating
black hole, we follow closely the Boyer-Lindquist spherical
coordinates $t,r,\theta,\varphi$ with the origin fixed on the central
black hole and the $z$-axis chosen as the axis of rotation.
Moreover, the self-gravity of the surrounding magnetofluid
is considered to be negligible in comparison with the gravitation of
the central object. Thus, the background geometry supporting the
disc is entirely determined by the central body and is defined by
Kerr metric
\begin{eqnarray}\label{Kerr Metric}
&&ds^2=\left(1-\frac{2 r}{\Sigma}\right) dt^2+\frac{4 a r\sin^2\theta}{\Sigma}\,dt\, d\varphi-\frac{\Sigma}{\Delta}\,dr^2\nonumber\\
&&\qquad-\Sigma\,d\theta^2-\frac{A\sin^2\theta}{\Sigma}\,
d\varphi^2,
\end{eqnarray}
where $\Delta=r^2-2 r+a^2$, $\Sigma=r^2+a^2\cos^2\theta$ and
$A=(r^2+a^2)^2-\Delta\, a^2 \sin^2\theta$. Throughout this paper, we
adopt geometric units $M=G=c=1$ as our basic scalings. Here, $M$,
$G$ and $c$ are the central black hole mass, the universal
gravitational constant, and the speed of light, respectively. This
implies $M=10 M_{\odot}$ as a unit of mass, and also $m=\frac{G
M}{c^2}$ and $t_0=\frac{G M}{c^3}$ as the units of length and time, respectively.
Furthermore, rotation of the black hole is parameterised by Kerr parameter $a$, as the total angular momentum per unit
mass of the black hole (i.e. $a=\frac{J}{Mc}$). Indeed, $a$ may be
measured in unit of length through a dimensionless spin parameter
$\alpha$ as $a=\alpha \, m$.

It is widely believed that black holes are probably maximally
rotating (Koide 2010; Tchekhovskoy, Narayan $\&$ McKinney 2011). However, even on a test particle level, solutions using
the fully rotating black hole seem to be an extremely formidable
task. To avoid this complexity, astrophysicists in analytic modelling tend to approximate
the black hole to be non-rotating characterised by the Schwarzschild
metric or to be slowly rotating characterised by the linearised Kerr
metric. However, in simulation community, this approximation is not popular (Porth et. al 2019).
Note that the slowly rotating regime which is an acceptable
approximation in the analytic astrophysics community means that one
considers up to linear order of the Kerr-rotating parameter in the
metric functions, governing equations and physical quantities
(Prasanna 1989; Rezzolla et al. 2001; Shaghaghian 2011; Harko, Kovacs $\&$ Lobo 2011).
Therefore, the linearised form of metric (\ref{Kerr Metric}) is
summarised as
\begin{eqnarray}\label{Linear Kerr Metric}
&&ds^2=\left(1-\frac{2}{r}\right)
dt^2-\left(1-\frac{2}{r}\right)^{-1} dr^2\nonumber\\
&&\qquad-r^2\left(d\theta^2+\sin^2\theta d\varphi^2\right)+ \frac{4
a}{r}\sin^2\theta \, dt\, d\varphi.\qquad
\end{eqnarray}

\subsubsection {Locally Non-Rotating Frame}

Once the background spacetime geometry rotates, it is necessary to
establish an inertial frame in which the frame-dragging effects of
the hole's spin are vanished. A set of local observers as zero
angular momentum observers are introduced (Yokosawa $\&$
Inui 2005). They rotate with the angular velocity $\omega$ and live
at constant $r$ and $\theta$, but at $\varphi=\omega t + const$.
This frame that becomes inertial at a far distance from the hole is
so-called locally non-rotating frame (LNRF). Applying slowly
rotating black hole approximation, the explicit transformations
between the LNRF and the Boyer-Lindquist frame given by Bardeen, Press, $\&$ Teukolsky (1972) are simplified as
\begin{eqnarray*}
\lambda^{(a)}_{\ \ i}=
\begin{bmatrix}
\left(1-\frac{2}{r}\right)^{1/2} & 0 & 0 & 0 \\
0 & \left(1-\frac{2}{r}\right)^{-1/2} & 0 & 0 \\
0 & 0 & r & 0 \\
-\frac{2 \,a }{r^2}\sin\theta & 0 & 0 & r\sin\theta
\end{bmatrix},
\end{eqnarray*}
and
\begin{eqnarray*}
\lambda^{i}_{\ (a)}=
\begin{bmatrix}
\left(1-\frac{2}{r}\right)^{-1/2} & 0 & 0 & 0 \\
0 & \left(1-\frac{2}{r}\right)^{1/2} & 0 & 0 \\
0 & 0 & \frac{1}{r} & 0 \\
\frac{2a }{r^3}\left(1-\frac{2}{r}\right)^{-1/2} & 0 & 0 &
\frac{1}{r\sin\theta}
\end{bmatrix},
\end{eqnarray*}
satisfying
$$\lambda^i_{(a)} \ \lambda^j_{(b)} \ g_{ij} = \eta_{(a)(b)},$$
where $g_{ij}$ and $\eta_{(a)(b)}$ are the metric and Minkowski
tensors, respectively. Noting here that parentheses around the
indices represent the components in LNRF. The physical variables are
transformed in this frame as follows:
\begin{eqnarray*}
F_{(\alpha)(\beta)}&=&\lambda^i_{\ (\alpha)}\  \lambda^j_{\ (\beta)}
\
F_{ij},\\
J^{(\alpha)}&=&\lambda^{(\alpha)}_{\ \ i} \  J^i,\\
V^{\alpha}&=& \frac{\lambda^{\alpha}_{\ (\beta)} \ V^{(\beta)}+
\lambda^{\alpha}_{\ (0)}} {\lambda^0_{\ (\beta)}\ V^{(\beta)}+
\lambda^0_{\ (0)}},
\end{eqnarray*}
where $F$ and $J$  are the electromagnetic field tensor and the 4-vector
electric current density, respectively. Moreover,
$V^{\alpha}$ is the spatial 3-velocity and is defined through
the relation $u^{\alpha}=
u^0 V^{\alpha}$ to the 4-velocity
$u$. We follow the (+,-,-,-) signature convention and the 4-velocity
normalisation condition as $u^i \, u_i=1$. It gives the following
general definition for the zeroth component of 4-velocity $u$
$$u^0=\left(g_{00}+2\, g_{0\alpha} V^{\alpha}+g_{\alpha \beta} V^{\alpha} V^{\beta}\right)^{-1/2},$$
which in linearised Kerr metric [equation (\ref{Linear Kerr Metric})] is
simplified as
\begin{eqnarray}\label{u0}
u^0=\left(1-\frac{2}{r}\right)^{-1/2} \left(1-V^2\right)^{-1/2}.
\end{eqnarray}
The total fluid velocity $V$ is related to its components as
$$V^2=\left[V^{(r)}\right]^2+\left[V^{(\theta)}\right]^2+\left[V^{(\varphi)}\right]^2,$$
wherein
\begin{eqnarray*}
&& V^{(r)} = \left(1-\frac{2}{r}\right)^{-1} V^r,\\
&& V^{(\theta)} = r \left(1-\frac{2}{r}\right)^{-1/2} V^{\theta},\\
&& V^{(\varphi)} = r \sin \theta
\left(1-\frac{2}{r}\right)^{-1/2}\left(V^{\varphi}-\frac{2\,
a}{r^3}\right).
\end{eqnarray*}
Furthermore, the components of field tensor and current density tensor are
transformed as
\begin{eqnarray}\label{transform}
&& B_r = r^2 \sin\theta B_{(r)},\nonumber\\
&& B_{\theta} = r \sin\theta \left(1-\frac{2}{r}\right)^{-1/2}
B_{(\theta)},\nonumber\\
&& B_{\varphi} = r \left(1-\frac{2}{r}\right)^{-1/2} B_{(\varphi)} ,\nonumber\\
&& E_r = \frac{2\, a}{r^2} \sin\theta
\left(1-\frac{2}{r}\right)^{-1/2} B_{(\theta)}+E_{(r)},\nonumber\\
&& E_{\theta}=-\frac{2 \, a}{r} \sin\theta \, B_{(r)}+r
\left(1-\frac{2}{r}\right)^{1/2} E_{(\theta)},\nonumber\\
&& E_{\varphi} = r \sin\theta \left(1-\frac{2}{r}\right)^{1/2}
E_{(\varphi)},\nonumber\\
&& J^r=\left(1-\frac{2}{r}\right)^{1/2} J^{(r)},\nonumber\\
&& J^{\theta}=\frac{1}{r}\, J^{(\theta)},\nonumber\\
&& J^{\varphi}=\frac{1}{r \sin\theta} J^{(\varphi)}+\frac{2\, a}{r^3} \left(1-\frac{2}{r}\right)^{-1/2} J^{(t)},\\
&& J^t=\left(1-\frac{2}{r}\right)^{-1/2} J^{(t)}\nonumber.
\end{eqnarray}

\subsubsection {Innermost Stable Circular Orbit}

\begin{figure}
\begin{center}
\includegraphics[width=7cm,height=7cm]{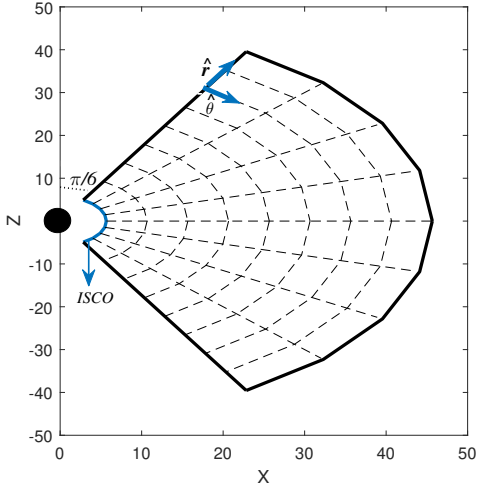}
\caption{Schematic sketch of the disc with the central black
hole.}\label{fig_Disc_config}
\end{center}
\end{figure}

The minimum allowed radius of charged particle trajectory that is
able to maintain stable circular orbit and do not enter into event
horizon of black hole is called innermost stable circular orbit (ISCO).
In the accretion disc theory, ISCO is regarded as one of the rotating
black hole's important features such as event horizon and ergosphere. ISCO is expected to be the inner edge of an accretion disc that
rotates around a black hole. The radius of ISCO is $6m$ in the case
of a Schwarzschild black hole, while for a Kerr black hole, it is
dependent on the black hole's spin by the following formula:
\begin{eqnarray*}
&& Z_1=1+(1-a^2)^{1/3} \left[\left(1+a\right)^{1/3}+(1-a)^{1/3}\right],\\
&& Z_2=\sqrt{3\, a^2+Z_1^2},\\
&& r_{ISCO}=3+Z_2-\sqrt{\left(3-Z_1\right)
\left(3+Z_1+2\,Z_2\right)}.
\end{eqnarray*}
In our model, we adopt the spherical coordinates of ($r,\, \theta,\,
\varphi$), defining the polar axis $\theta=0$ and $\theta=\pi$ to be
perpendicular to the disc plane. We set a computational domain of
$r_{ISCO} \leq r \leq 50 m$ and $\frac{\pi}{6}\leq \theta \leq
\pi-\frac{\pi}{6}$. We draw the schematic depiction of our disc in
Figure \ref{fig_Disc_config}. It shows that the meridional structure of the
disc extends to about $\frac{\pi}{3}$ 
on either side of the equatorial
plane. These ranges for the radius $r$ and angular thickness of the
disc are assumed typically. Although, as we will see later, some
physical circumstances will vary these intervals.

\subsubsection {Keplerian Velocity Distribution}

In Newtonian gravity, angular momentum $l^*$ and angular velocity
$\Omega$ are related by the formula $l^*=r^2 \Omega$, and therefore,
there is no ambiguity in defining a non-rotating frame as
$\Omega=0=l^*$. However, in the rotating Kerr geometry $l^* \propto
(\Omega-\omega)$, wherein $\Omega=\frac{d \varphi}{dt}$ is the
angular velocity of the orbiting matter and
$\omega=-\frac{g_{t\varphi}}{g_{\varphi\varphi}}=\frac{2 a}{r^3}$ is
that of the frame dragging of the LNRF relative to distant
observers. Generally, the azimuthal component of the 3-velocity in
the LNRF reads $V^{(\varphi)}=\frac{d x^{(\varphi)}}{d x^{(0)}}$,
where $x^{(0)}$ and $x^{(\varphi)}$ are the time and spatial
coordinates in the LNRF, respectively,
\begin{eqnarray*}
&& d x^{(0)}=\lambda^{(0)}_{\ i} d x^i=\left(1-\frac{2}{r}\right)^{1/2} dt,\\
&& d x^{(\varphi)}=\lambda^{(\varphi)}_{\ i} d x^i=r\sin\theta \,
\left[d\varphi-\omega \, dt\right].
\end{eqnarray*}
Afterwards
$$V^{(\varphi)}=r \sin\theta \left(1-\frac{2}{r}\right)^{-1/2} \left[\Omega-\omega\right].$$
If the matter follows nearly circular orbits characterised by the
Keplerian distributions of the angular velocity
$$\Omega_{K}=\frac{1}{r^{3/2}+a},$$
then the azimuthal component of the Keplerian 3-velocity in the LNRF
is obtained as
$$V^{(\varphi)}_K=r \sin\theta \left(1-\frac{2}{r}\right)^{-1/2} \left(\frac{1}{r^{3/2}+a}-\frac{2\, a}{r^3}\right).$$

\subsection {Seed Magnetic field Model}

As a matter of fact, in problem of magnetised accretion discs, the
magnetic field of a resistive disc is not totally arisen from the
electric current of the plasma disc. Strength of the disc current is
determined in agreement with that of the external field as well. In
such a situation, there is always an external field penetrating the
disc (Kaburaki 1987). We describe the magnetic field as a
superposition of the seed field $\textbf{B}^S$ caused by some
external sources and the disc field $\textbf{B}^D$ induced by the
current flowing in the disc
$$\textbf{B}=\textbf{B}^S+\textbf{B}^D.$$
Roughly speaking, magnetosphere develops well in the place where the
strength of the seed field exceeds that of the disc field (i.e.
$\left|\textbf{B}^S\right|\geq\left|\textbf{B}^D\right|$). And also,
magnetodisc is the region where
$\left|\textbf{B}^S\right|<<\left|\textbf{B}^D\right|$. Thus within
the disc, $\textbf{B}$ can be replaced by $\textbf{B}^D$, with a
good accuracy. In this way, the magnetic field appearing in the next
sections is the same as $\textbf{B}^D$, that its superscript $D$ has
been dropped for simplicity. From now on, we put superscript just
for the seed field $\textbf{B}^S$.

Realistic cases of a rotating black hole with a disc and magnetic
field are likely to be extremely complicated. However, some authors
have discussed the magnetic field associated with the black hole as
a somewhat poloidal structure and have modelled it as being generated
by some toroidal electric current rings exterior to the black hole's
event horizon (Li 2000; Ghosh 2000; Tomimatsu $\&$ Takahashi 2001).
We apply this poloidal structure as a dipolar model for the seed
magnetic field (Prasanna $\&$ Vishveshwara 1978; Takahashi $\&$
Koyama 2009)
\begin{eqnarray*}
&& B_{(r)}^S =-\frac{3\mu}{4\gamma^2} \,
\frac{2\cos\theta}{r^2\Sigma}\left\{W
\left(1+\frac{a^2 \sin^2\theta}{\Sigma}\right)\right.\\
&& \qquad\quad \left.-\sin^2\theta\left[\left(r-1\right) a^2-\frac{\Delta
a^2}{2\gamma}\ln\left(\frac{r-r_-}{r-r_+}\right)\right]\right\},\\[.3cm]
&& B_{(\theta)}^S=\frac{3\mu}{4\gamma^2}
\left(1-\frac{2}{r}\right)^{1/2}\frac{1}{r \Sigma} \left(\frac{-2r
W}{\Sigma}+\frac{\partial W}{\partial r}\right) \sin\theta.
\end{eqnarray*}
Here, $\mu$ is the magnetic dipole moment of the central black hole
which relates to its surface magnetic field $B_s$ and radius $R$ as
$\mu=B_s R^3$. We take $\mu=1260 \times 10^{27} \ Gauss.cm^3$, that is, an
appropriate choice for the intrinsic magnetic moment of a typical
$10 M_{\odot}$ black hole (Robertson $\&$ Leiter 2002). It is worth
noting that $\Sigma$ and $\Delta$ have the previous definitions and
the other undefined variables are
\begin{eqnarray*}
r_{\pm}&=& 1\pm\gamma,\\
\gamma &=&\sqrt{1-a^2},\\
Q &=&\frac{1}{2\gamma}\ln\left(\frac{r-r_-}{r-r_+}\right),\\
W &=& \left(r-1\right) a^2 \cos^2\theta+r \left(r^2+r+2 a^2\right)\\
&-&\left[r\left(r^3-2 a^2+a^2 r\right)+\Delta \,
a^2\cos^2\theta\right] Q,\\
\frac{\partial W}{\partial r} &=& a^2 \cos^2\theta+3 r^2+2 r+2 a^2\\
&-&\left[4 r^3-2 a^2+2 a^2 r+2 a^2 \cos^2\theta
(r-1)\right]Q\\
&+&\left[r(r^3-2 a^2+a^2 r)+\Delta a^2
\cos^2\theta\right]\frac{1}{(r-1)^2-\gamma^2}.
\end{eqnarray*}

\begin{figure}
\begin{center}
\includegraphics[width=7cm,height=7cm]{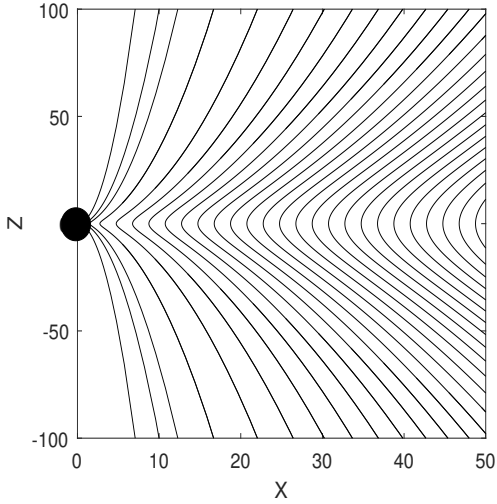}
\caption{Dipolar magnetic filed model of
the central black hole in the meridional plane.}\label{magfield_BH}
\end{center}
\end{figure}

Figure \ref{magfield_BH} shows a typical profile of the dipolar
magnetic filed structure of central black hole at infinity (Appendix
A).

\section{General Formalism}

\subsection{Basic Equation}
Fully general relativistic MHD equations governing
the motion of the resistive magnetised plasma accreted by a central
compact object are mass conservation or continuity equation
\begin{eqnarray}\label{Consv rho u}
(\rho_0 u^i)_{\, ;i}=0,
\end{eqnarray}
and energy - momentum conservation law
\begin{eqnarray}\label{Consv Tij}
T^{ij}_{\ ;j}=0,
\end{eqnarray}
supplemented by Maxwell equations
\begin{eqnarray}
\label{Max Equ1} &&F^{ij}_{\ ;j}=-4\pi J^i,\\[.3cm]
\label{Max Equ2} &&F_{ij,k}+F_{ki,j}+F_{jk,i}=0,
\end{eqnarray}
and the generalised Ohm's law
\begin{eqnarray}\label{Ohm law}
J^i=\sigma F^i_{\ k} u^k,
\end{eqnarray}
wherein $\sigma$ is the electric conductivity which is assumed
constant for simplicity. It is worth noting here that semicolon
stands for covariant derivative and comma for partial derivative.
Latin indices denote spacetime components (0-3) and Greek ones
denote spatial components (1-3). Furthermore, we adopt the standard
convention for the summation over the repeated indices.

Our MHD system will be specified by the following
choice for the energy-momentum tensor
$$T^{ij}=T^{ij}_{Fluid}+T^{ij}_{Em}.$$
It consists of a fluid part
\begin{eqnarray*}
T^{ij}_{Fluid}=\left(\rho+p \right) u^i u^j-p\ g^{ij},
\end{eqnarray*}
and an electromagnetic part
\begin{eqnarray*}
T^{ij}_{Em}=-\frac{1}{4\pi}\left(F^{ik} \ F^j_{\ k}-\frac{1}{4}\
g^{ij}\ F_{kl}\ F^{kl}\right),
\end{eqnarray*}
where $p$ is the gas pressure and $\rho=\rho_0+u$ is the total density of mass-energy
including the rest mass density $\rho_0$ and the internal energy per unit volume $u$.
On the other hand, the electromagnetic field tensor is related to
the electric and magnetic fields through
$$E_{\alpha}=F_{\alpha t},\quad\quad B_{\alpha}=\epsilon_{\alpha \beta \gamma} F_{\beta \gamma},$$
where $\epsilon_{\alpha \beta \gamma}$ is the Levi-Civita symbol.
For an axisymmetric and stationary
magnetofluid disc, all flow variables that neither depend on the time $t$
nor on the azimuthal coordinate $\varphi$ are functions of only $r$
and $\theta$. Then, equations (\ref{Max Equ1}) and (\ref{Max
Equ2}) may be expanded
\begin{eqnarray}\label{Max1}
4\pi
J^r=-\frac{1}{r^2\sin\theta}\left(1-\frac{2}{r}\right)\frac{\partial}{\partial\theta}\left(\sin\theta
B_{\varphi}\right),
\end{eqnarray}
\begin{eqnarray}\label{Max2}
4\pi J^{\theta}=\frac{1}{r^2}\frac{\partial}{\partial r
}\left[\left(1-\frac{2}{r}\right)B_{\varphi}\right],
\end{eqnarray}
\begin{eqnarray}\label{Max3}
4\pi J^{\varphi}=\frac{1}{r^4
\sin\theta}\frac{\partial}{\partial\theta}\left(\frac{B_r}{\sin\theta}\right)
-\frac{1}{r^2 \sin^2\theta} \frac{\partial}{\partial r
}\left[\left(1-\frac{2}{r}\right) B_{\theta}\right]\nonumber\\[.3cm]
-\frac{2 a}{r^2}\left[\frac{\partial}{\partial
r}\left(\frac{E_r}{r}\right)+\frac{1}{r^3\sin\theta}\left(1-\frac{2}{r}\right)^{-1}
\frac{\partial}{\partial\theta}\left(\sin\theta
E_{\theta}\right)\right],
\end{eqnarray}
\begin{eqnarray}\label{Max4}
4\pi J^t=-\frac{1}{r^2} \frac{\partial}{\partial r}\left(r^2
E_r\right)-\frac{1}{r^2
\sin\theta}\left(1-\frac{2}{r}\right)^{-1}\frac{\partial}{\partial\theta}\left(\sin\theta
E_{\theta}\right)\nonumber\\[.3cm]
-\frac{2 a}{r^5
\sin\theta}\left(1-\frac{2}{r}\right)^{-1}\frac{\partial}{\partial\theta}\left(\sin\theta
B_r\right)+\frac{2 a}{r^2}\frac{\partial}{\partial r
}\left(\frac{B_{\theta}}{r}\right),
\end{eqnarray}
\begin{eqnarray}\label{Max5}
\frac{\partial B_{\theta}}{\partial \theta}+\frac{\partial
B_r}{\partial r}=0,
\end{eqnarray}
\begin{eqnarray}\label{Max6}
\frac{\partial E_{\theta}}{\partial r}-\frac{\partial E_r}{\partial
\theta}=0,
\end{eqnarray}
\begin{eqnarray}\label{Max7}
\frac{\partial E_{\varphi}}{\partial r}=0,
\end{eqnarray}
\begin{eqnarray}\label{Max8}
\frac{\partial E_{\varphi}}{\partial \theta}=0.
\end{eqnarray}

\subsection{Possible Solutions}
\subsubsection{Simplifying Assumptions}
Equations (\ref{Max7}) and (\ref{Max8}) state that the toroidal electric field must be constant.
This constant value may be chosen equal to zero for simplicity.
Now, we need a wise assumption
to simplify the highly non-linear coupled equations (\ref{Ohm
law}) - (\ref{Max6}). If we assume
\begin{eqnarray}\label{B_phi}
B_{\varphi}=\frac{b_{\varphi}}{\sin\theta}\left(1-\frac{2}{r}\right)^{-1},
\end{eqnarray}
wherein $b_{\varphi}$ is constant, then $J^r=J^{\theta}=0$ through equations (\ref{Max1}) and (\ref{Max2}).
Also through the Ohm's law (\ref{Ohm law}), it results in
\begin{eqnarray*}
&& E_r = B_{\theta} \, V^{\varphi} - B_{\varphi} \, V^{\theta},\\
&& E_{\theta} = -B_r \, V^{\varphi} + B_{\varphi} \, V^r,
\end{eqnarray*}
that keep the same forms in LNRF as
\begin{eqnarray}
\label{E&B1}
&& E_{(r)} = B_{(\theta)} \, V^{(\varphi)} - B_{(\varphi)} \, V^{(\theta)},\\
\label{E&B2}
&& E_{(\theta)} = -B_{(r)} \, V^{(\varphi)} + B_{(\varphi)} \, V^{(r)}.
\end{eqnarray}
The other non-zero components of the Ohm's law (\ref{Ohm law}) gives
\begin{eqnarray*}
J^{\varphi}=-\sigma \left(B_{\theta} u^r- B_r u^{\theta} \right)
\left[\frac{1}{r^2 \sin^2\theta}+\frac{2\,
a}{r^3}\left(1-\frac{2}{r}\right)^{-1} V^{\varphi}\right],
\end{eqnarray*}
\begin{eqnarray*}
J^t&=&-\sigma \left(1-\frac{2}{r}\right)^{-1} \left(B_{\theta} u^r-
B_r u^{\theta} \right)\left(V^{\varphi}-\frac{2 \, a}{r^3}\right)\\
&=&-\frac{\sigma}{r\sin\theta} \left(1-\frac{2}{r}\right)^{-1/2}
\left(B_{\theta} u^r- B_r u^{\theta} \right) V^{(\varphi)},
\end{eqnarray*}
that they find a more concise form, with the help of transformations
(\ref{transform})
\begin{eqnarray}
\label{J_phi_Ohm}
&& J^{(\varphi)}=-\sigma u^0 \left(1-\frac{2}{r}\right)^{1/2} \left[B_{(\theta)} V^{(r)}-B_{(r)} V^{(\theta)}\right],\qquad\\
\label{J_t_Ohm}
&& J^{(t)}=J^{(\varphi)} V^{(\varphi)}.
\end{eqnarray}
Defining the total derivative
$$D \equiv V^{(r)}\frac{\partial}{\partial r}+\left(1-\frac{2}{r}\right)^{-1/2}\frac{V^{(\theta)}}{r}\frac{\partial}{\partial \theta},$$
and combining equations  (\ref{Consv rho u}) and the zeroth
component of equation (\ref{Consv Tij}), continuity equation is
achieved in LNRF
\begin{eqnarray}\label{Continuity equ1}
&&\left(\rho+p\right) \left\{\frac{\partial V^{(r)}}{\partial
r}+\frac{2}{r}V^{(r)}+\frac{1}{r}\left(1-\frac{2}{r}\right)^{-1/2}
\left[\frac{\partial V^{(\theta)}}{\partial \theta}\right.\right.\nonumber\\[.1cm]
&&\left.\left.+\cot\theta V^{(\theta)}\right]+\frac{6 a}{r^3} \left(1-\frac{2}{r}\right)^{-1/2} \sin \theta \, V^{(r)} \, V^{(\varphi)} \right\}\nonumber\\[.1cm]
&&+D \left(\rho-p\right)=\nonumber\\
&&\frac{2}{\sigma u^0} \left(1-\frac{2}{r}\right)^{-1}
\left[J^{(\varphi)}\right]^2
\left\{1-\left[V^{(\varphi)}\right]^2\right\},
\end{eqnarray}
and also the momentum equations are obtained from the spatial
components of equation (\ref{Consv Tij})
\begin{eqnarray}\label{Radial equ1}
&&\left(\rho+p\right)\left(1-V^2\right)^{-1}
\left[D V^{(r)}-\frac{1}{r}\left\{[V^{(\theta)}]^2+[V^{(\varphi)}]^2\right\}\right.\nonumber\\[.1cm]
&&\left.+\frac{1}{r^2}\left(1-\frac{2}{r}\right)^{-1}\left\{1-\left[V^{(r)}\right]^2\right\}\right.\nonumber\\[.1cm]
&&\left.-\frac{6\, a}{r^3}\left(1-\frac{2}{r}\right)^{-1/2}
\sin\theta \, V^{(\varphi)}
\left\{1-\left[V^{(r)}\right]^2\right\}\right]
+\frac{\partial p}{\partial r}\nonumber\\[.1cm]
&&-\left(1-\frac{2}{r}\right)^{-1/2} B_{(\theta)} J^{(\varphi)} \,\left\{1-\left[V^{(\varphi)}\right]^2\right\}=0,\qquad
\end{eqnarray}

\begin{eqnarray}\label{Meridional equ1}
&&\left(\rho+p\right)\left(1-V^2\right)^{-1}
\left\{D V^{(\theta)}+\frac{\left(1-\frac{3}{r}\right)}{\left(1-\frac{2}{r}\right)}\frac{V^{(r)}V^{(\theta)}}{r}\right.\nonumber\\
&&\left.+\frac{6\, a}{r^3}\left(1-\frac{2}{r}\right)^{-1/2}
\sin\theta \, V^{(r)}\, V^{(\theta)}\, V^{(\varphi)}\right.\nonumber\\[.1cm]
&&\left.-\cot\theta\left(1-\frac{2}{r}\right)^{-1/2}\frac{\left[V^{(\varphi)}\right]^2}{r}\right\}+\left(1-\frac{2}{r}\right)^{-1/2}
\frac{1}{r}\frac{\partial p}{\partial\theta}\nonumber\\[.1cm]
&&+\left(1-\frac{2}{r}\right)^{-1/2}B_{(r)}\,
J^{(\varphi)}\left\{1-\left[V^{(\varphi)}\right]^2\right\}=0,
\end{eqnarray}

\begin{eqnarray}\label{Azimuthal equ1}
&&DV^{(\varphi)}+\left(1-\frac{2}{r}\right)^{-1}\left(1-\frac{3}{r}\right)\frac{V^{(r)}V^{(\varphi)}}{r}\nonumber\\[.1cm]
&&+\cot\theta\left(1-\frac{2}{r}\right)^{-1/2}\frac{V^{(\theta)}V^{(\varphi)}}{r}\nonumber\\[.1cm]
&&+\frac{6\, a}{r^3}\left(1-\frac{2}{r}\right)^{-1/2} \sin\theta \,
V^{(r)} \left[V^{(\varphi)}\right]^2=0.\qquad
\end{eqnarray}
Equation (\ref{Azimuthal equ1}) may be summarised as
\begin{eqnarray}\label{Azimuthal equ2}
D \widetilde{V}^{(\varphi)} = - \frac{6\, a}{r^4}\, V^{(r)}
\left[\widetilde{V}^{(\varphi)}\right]^2,
\end{eqnarray}
while we define a new variable
$$\widetilde{V}^{(\varphi)}=r \sin\theta \left(1-\frac{2}{r}\right)^{-1/2} V^{(\varphi)}.$$
To have an integrable form for equation (\ref{Azimuthal equ2}), we
multiply both its sides, by an integration constant $L$,
$$L \frac{D \widetilde{V}^{(\varphi)}}{\left[\widetilde{V}^{(\varphi)}\right]^2}= -D \left(1-\frac{2 \, a \,L}{r^3}\right).$$
Now, it can integrate simply as
$$\widetilde{V}^{(\varphi)}=L \left(1-\frac{2 \, a \, L}{r^3}\right)^{-1}.$$
Indeed, due to the dimensional considerations, $L$ is defined as
$L=l\, m \, c$,  wherein $l$ is called the angular momentum
parameter. Ultimately, the final solution for the azimuthal velocity
is obtained
\begin{eqnarray}\label{V_f}
V^{(\varphi)}=\frac{L}{r \sin\theta}
\left(1-\frac{2}{r}\right)^{1/2} \left(1-\frac{2 \, a
\,L}{r^3}\right)^{-1}.
\end{eqnarray}
Substituting equations (\ref{J_t_Ohm}) and (\ref{V_f}), the
transformation equation (\ref{transform}) for $J^{\varphi}$ gets a
shorter form as
$$J^{\varphi}=\frac{1}{r \sin\theta} \left(1-\frac{2 \, a \, L}{r^3}\right)^{-1} J^{(\varphi)}.$$ To brief the appearance of the
governing equations (\ref{Continuity equ1}) - (\ref{Azimuthal
equ1}), we multiply equation (\ref{Radial equ1}) by $V^{(r)}$,
equation (\ref{Meridional equ1}) by $V^{(\theta)}$ and equation
(\ref{Azimuthal equ1}) by $V^{(\varphi)}$ and adding
\begin{eqnarray}\label{Vr*Radial+Vt*Meridional1}
&&\left(\rho+p\right)D \ln
\left[\left(1-\frac{2}{r}\right)\left(1-\frac{2 \, a \,L}{r^3}\right)^{-2}\left(1-V^2\right)^{-1}\right]\nonumber\\
&&+2\ D\, p =-\frac{2}{\sigma
u^0}\left(1-\frac{2}{r}\right)^{-1}\left[J^{(\varphi)}\right]^2\left\{1-[V^{(\varphi)}]^2\right\}.\quad\
\end{eqnarray}
Continuity equation (\ref{Continuity equ1}) can be simplified too
\begin{eqnarray}\label{Continuity equ2}
&&\left(\rho+p\right) \left[{\nabla \cdot \mathbf{\tilde{V}}}+D \ln
\left(1-\frac{2 \, a \,L}{r^3}\right)\right]+D \left(\rho-p\right)\nonumber\\
&&\qquad=\frac{2}{\sigma u^0}\left(1-\frac{2}{r}\right)^{-1}
\left[J^{(\varphi)}\right]^2\left\{1-[V^{(\varphi)}]^2\right\},
\end{eqnarray}
with a new definition for total fluid velocity as
\begin{equation}
\mathbf{\tilde{V}}=V^{(r)}\ \hat{r}+\tilde{V}^{(\theta)}\
\hat{\theta}+V^{(\varphi)}\ \hat{\varphi},
\end{equation}
wherein
$\tilde{V}^{(\theta)}=V^{(\theta)}\left(1-\frac{2}{r}\right)^{-1/2}$.
As seen, right-hand side of these latter two equations
(\ref{Vr*Radial+Vt*Meridional1}) and (\ref{Continuity equ2}) are
similar with opposite sign. It motivates us to add them to get rid of
this long term
\begin{eqnarray}\label{Adding equ1}
{\nabla \cdot \mathbf{\tilde{V}}}+D
\ln\left[\frac{\left(\rho+p\right)\left(1-\frac{2}{r}\right)}{\left(1-\frac{2
\, a \,L}{r^3}\right) \left(1-V^2\right)}\right]=0.
\end{eqnarray}
Therefore, we have summarised the motion equations (\ref{Continuity
equ1}) - (\ref{Azimuthal equ1}) in an equation (\ref{Adding equ1}).
To achieve an integrable form for it, we must try to write the term
${\nabla \cdot \mathbf{\tilde{V}}}$ in terms of the total derivative
$D$. This term may be written as
$${\nabla \cdot \mathbf{\tilde{V}}}=D \ln\left(r^2 \sin\theta \ V^{(r)}\right)+\frac{\tilde{V}^{(\theta)}}{r}\frac{\partial}{\partial \theta}\ln
\left(\frac{\tilde{V}^{(\theta)}}{V^{(r)}}\right).$$ In order to
reach to an integrable form, we try to express the second term in
terms of $D$. To this aim, one may assume that the poloidal
component of total fluid velocity including $V^{(r)}$ and
$\tilde{V}^{(\theta)}$ are two separable functions of their
independent variables $r$ and $\theta$ as $V^{(r)}=V_1^{(r)}(r) \
V_2^{(r)} (\theta)$ and $\tilde{V}^{(\theta)}=V_1^{(\theta)}(r) \
V_2^{(\theta)} (\theta)$, respectively. Now, if their radial dependencies are
presumed to be similar [i.e. $V_1^{(r)}(r)=V_1^{(\theta)}(r)$], then
the term $\frac{\tilde{V}^{(\theta)}}{V^{(r)}}$ is a function only
of $\theta$ as
\begin{eqnarray}\label{assum_Vr}
\frac{\tilde{V}^{(\theta)}}{V^{(r)}}=\frac{1}{C_1(\theta)},
\end{eqnarray}
and equation (\ref{Adding equ1}) is rewritten as
\begin{eqnarray}\label{Adding equ2}
D\ln\left[\frac{\left(\rho+p\right)\left(1-\frac{2}{r}\right)}{\left(1-\frac{2
\, a \,L}{r^3}\right) \left(1-V^2\right)}\, r^2 \sin\theta
V^{(r)}\right]-D \ln C_1(\theta)=0.
\end{eqnarray}
In fact, the term in bracket can be interpreted as mass accretion
rate (Shaghaghian 2016)
\begin{eqnarray}\label{mass accretion1}
\dot{M}=\frac{\left(\rho+p\right)\left(1-\frac{2}{r}\right)}{\left(1-\frac{2
\, a \,L}{r^3}\right) \left(1-V^2\right)}\, r^2 \sin\theta V^{(r)}.
\end{eqnarray}
Thus, equation (\ref{Adding equ2}) leads to
\begin{eqnarray}\label{mass accretion2}
\dot{M}=\dot{M}_0 C_1(\theta),
\end{eqnarray}
wherein, $\dot{M}_0$ is an integration constant and $C_1(\theta)$ is
an unknown function that will be determined. We prefer a
sub-Eddington regime. Since as elucidates in the introduction, the
super-Eddington accretion discs are generally expected to possess
vortex funnels and radiation pressure driven jets (Okuda 2002).
Although this aspect is so noteworthy in recent decades, it is
beyond the scope of this paper and must be pursued separately. Thus,
we choose $\dot{M}_0 = - 10^{-8} \frac{M_{\odot}}{year} $, which is
a normal mass accretion rate for a typical $M = 10\, M_{\odot}$
black hole (Koide 2010).

\subsubsection{Disc Magnetic Field Model}

Now, it is time to return to the remaining Maxwell equations
(\ref{Max3}) - (\ref{Max6}) and rewrite them in LNRF with the aid
of transformation equations (\ref{transform}),
\begin{eqnarray}\label{Max3_Jphi}
&&4\pi J^{(\varphi)}=\frac{1}{r} \left(1-\frac{2 \, a
\,L}{r^3}\right)\left[\frac{\partial B_{(r)}}{\partial
\theta}-\frac{\partial}{\partial r}\left[r
\left(1-\frac{2}{r}\right)^{1/2} B_{(\theta)}\right]\right.\nonumber\\
&&\left.-\frac {2\, a}{r}\left\{r\sin\theta \frac{\partial}{\partial
r}\left[\frac{E_{(r)}}{r}\right]\right.\right.\nonumber\\
&&\left.\left.+\frac{1}{r}\left(1-\frac{2}{r}\right)^{-1/2}\frac{\partial}{\partial
\theta}\left[\sin\theta E_{(\theta)}\right]\right\}\right],
\end{eqnarray}
\begin{eqnarray}\label{Max4_Jt}
4\pi J^{(t)}=\frac{-1}{r^2}\left(1-\frac{2}{r}\right)^{1/2}
\frac{\partial}{\partial}\left[r^2
E_{(r)}\right]-\frac{1}{r\sin\theta}\frac{\partial}{\partial
\theta}\left[\sin\theta E_{(\theta)}\right].
\end{eqnarray}

\begin{eqnarray}\label{Max5_1}
\sin\theta\frac{\partial}{\partial r}\left[r^2 B_{(r)}\right]+r
\left(1-\frac{2}{r}\right)^{-1/2}\frac{\partial}{\partial\theta}\left[\sin\theta
\ B_{(\theta)}\right]=0,
\end{eqnarray}
\begin{eqnarray}\label{Max6_1}
\frac{\partial}{\partial r}\left[A \,
B_{(r)}\right]+\frac{1}{r}\left(1-\frac{2}{r}\right)^{-1/2}\frac{\partial}{\partial
\theta}\left[A \, B_{(\theta)}\right]=0,
\end{eqnarray}
where
$$A=\frac{2\, a}{r} \sin\theta + r \left(1-\frac{2}{r}\right)^{1/2} V^{(\varphi)}.$$
As expected, due to the relations (\ref{E&B1}), (\ref{E&B2}), and (\ref{J_t_Ohm}),
equations (\ref{Max3_Jphi}) and (\ref{Max4_Jt}) are not
independent and achieve a similar appearance
\begin{eqnarray}\label{Max3_(Jphi)}
&&4\pi J^{(\varphi)}=\frac{1}{r}\frac{\partial B_{(r)}}{\partial
\theta}-\frac{1}{r}\frac{\partial}{\partial r}\left[r
\left(1-\frac{2}{r}\right)^{1/2} B_{(\theta)}\right]\nonumber\\
&&+\frac{6 a L}{r^4} \left(1-\frac{2}{r}\right)^{1/2} \left(1-\frac{2 \, a
\,L}{r^3}\right)^{-1} B_{(\theta)}\nonumber\\
&&+\frac{2 a}{r^2} \left(1-\frac{2 \, a \,L}{r^3}\right)\left\{r \sin\theta
\frac{\partial}{\partial r}\left[B_{(\varphi)}\frac{V^{(\theta)}}{r}\right]\right.\nonumber\\
&&\left.-\frac{1}{r} \left(1-\frac{2}{r}\right)^{-1/2} \frac{\partial}{\partial \theta}\left[\sin\theta B_{(\varphi)} V^{(r)}\right]\right\}.
\end{eqnarray}
$b_{\varphi}$ as a free parameter is chosen small, so that the last term in $J^{(\varphi)}$ [equation (\ref{Max3_(Jphi)})] may be ignorable. Because it is the product of two small parameters $a$ and $b_{\varphi}$.

\begin{figure}
\begin{center}
\includegraphics[width=7cm,height=7cm]{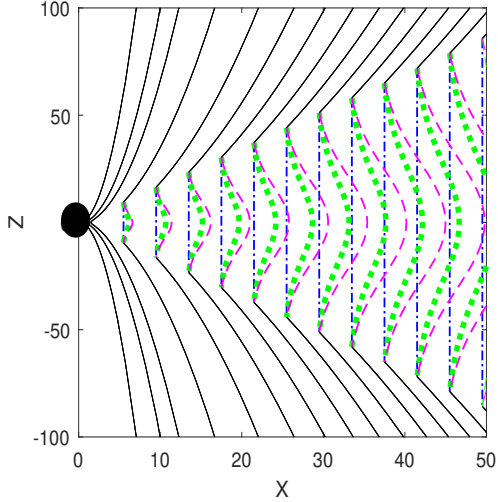}
\caption{Structure of magnetic field
lines in presence of disc field projected on the meridional plane
of the disc. The solid lines being the same as Figure
\ref{magfield_BH} represent the dipolar magnetic filed of the
central black hole, while the dashed lines represent the disc's
field in the case $b_{\varphi}=0$. The different colours correspond to the different spin
parameter $a$: blue -. is $a=0$, green dotted is $a=0.1$ and magenta
$--$ denotes $a=0.2$.}\label{magfield_Disc}
\end{center}
\end{figure}

Poloidal magnetic field of the disc is usually achieved
through the self-consistent solution of equations (\ref{Max5_1})
and (\ref{Max6_1}). This is what happens in the case of
Schwarzschild metric (Shaghaghian 2016). However, in the case of
Kerr metric, by virtue of the presence of the term $\frac{2\, a}{r}
\sin\theta $ in $A$, the self-consistent solution that satisfies
these two equations simultaneously is not possible. Consequently,
to find the poloidal magnetic field, they have to be solved
separately (Shaghaghian 2011). As a result, we go to equation
(\ref{Max5_1}) which seems to be easier to solve. To this aim, we
presume the following model for the poloidal component of the disc's
magnetic field
\begin{eqnarray*}
&& B_{(r)}=b_1 (r)\ \cot\theta \ b_2(\theta),\\
&& B_{(\theta)}=f(r)\  b_1(r)\  b_2(\theta),
\end{eqnarray*}
here
$$b_1(r)= - B_1 r^{k-2} \left(1-\frac{2}{r}\right)^{-k/2} \left(1-\frac{2 \, a \, L}{r^3}\right)^k,$$ wherein $k$ and $B_1$ are constants and $b_2(\theta)$
and $f(r)$ are the unknown functions that must be determined. It is
valuable to mention that, this model is inspired us by the
self-consistent solution for poloidal magnetic field in the
Schwarzschild metric (Shaghaghian 2016). Substituting this model in
equation (\ref{Max5_1}), we have
\begin{eqnarray*}
\frac{\left(1-\frac{2}{r}\right)^{1/2}}{r\, f(r)\,
b_1(r)}\frac{d}{dr}\left[r^2 b_1(r)\right]+\frac{1}{b_2(\theta)\,
\cos\theta} \frac{d}{d\theta}\left[\sin\theta \,
b_2(\theta)\right]=0.
\end{eqnarray*}
As seen, functions of the variables $r$ and $\theta$ have been
separated. Thus, each part must be constant
\begin{eqnarray*}
&& \frac{d \left[\sin\theta \, b_2(\theta)\right]}{b_2(\theta)} = k
\,
\cos\theta \, d\theta,\\
&& f(r)=\frac{1}{-k \, r \, b_1(r)} \left(1-\frac{2}{r}\right)^{1/2}
\frac{d}{dr} \left[r^2 b_1(r)\right].
\end{eqnarray*}
Solving these two equations, the unknown functions in our model are
obtained
\begin{eqnarray*}
b_2 (\theta) &=& \sin^{k-1} \theta,\\
f(r) &=& -\left(1-\frac{2}{r}\right)^{-1/2} A_1(r),
\end{eqnarray*}
wherein
$$A_1(r)=\left[\left(1-\frac{3}{r}\right)+\frac{6 \, a \,
L}{r^3} \left(1-\frac{2}{r}\right) \left(1-\frac{2 \, a
\,L}{r^3}\right)^{-1} \right].$$
Then, the components of the poloidal magnetic field are achieved
\begin{eqnarray}\label{Br}
B_{(r)}=-B_1 \, r^{k-2} \, \frac{\left(1-\frac{2 \, a
\,L}{r^3}\right)^k}{\left(1-\frac{2}{r}\right)^{k/2}} \, \sin^{k-2}
\theta \, \cos\theta,
\end{eqnarray}
\begin{eqnarray}\label{Bt}
B_{(\theta)}=B_1 \, r^{k-2} \, \frac{\left(1-\frac{2 \, a
\,L}{r^3}\right)^k}{\left(1-\frac{2}{r}\right)^{(k+1)/2}} \, A_1(r)
\, \sin^{k-1} \theta.
\end{eqnarray}
$B_1$ is a definite constant that may be found as a result of
continuity of the magnetic field lines across the disc boundary
surface
\begin{eqnarray}\label{matching condition}
\left(B^D\right)^2|_{r=r_0,\
\theta=\frac{\pi}{6}}=\left(B^S\right)^2|_{r=r_0,\
\theta=\frac{\pi}{6}},
\end{eqnarray}
where
\begin{eqnarray*}
&&\left(B^D\right)^2=B_{(r)}^2+B_{(\theta)}^2,\\[.2cm]
&&\left(B^S\right)^2=\left[B^S_{(r)}\right]^2+\left[B^S_{(\theta)}\right]^2,
\end{eqnarray*}
and $r_0$ is the radius where two field lines connect together.
Ghosh $\&$ Lamb (1979a, b) notified that the external magnetic field
penetrates the disc via a variety of processes owing to the presence
of a finite resistivity. In fact, electrical conductivity is treated
as a measure of the rate of field line slippage through the plasma
disc. Figure \ref{magfield_Disc} shows the magnetic field lines of the
disc connected with the undistorted dipolar magnetic field lines of
the central hole at the surface of the disc. Additionally, it is
evident that the magnetic filed lines inside the disc are pushed
outwards. As black hole spins faster, this outward push becomes more
(Appendix A).

Now, we profit this occasion and define the magnetic pressure both
in the disc as $p_{mag}^D=\frac{\left(B^D\right)^2}{8 \pi}$ and
within the magnetosphere surrounding the central black hole as
$p_{mag}^S=\frac{\left(B^S\right)^2}{8 \pi}$. We study the effect of
disc magnetic pressure via a new physical variable $\beta$ defined
as the ratio of the gas pressure to the magnetic pressure in the
disc. Thus, $\beta=\frac{p}{p_{mag}^D}$.

\subsubsection{Analytical and Numerical Solutions}

Substituting the equations (\ref{Br}) and (\ref{Bt}) in equation
(\ref{Max3_(Jphi)}), the azimuthal current density
is obtained
\begin{eqnarray}\label{J_phi_Max}
J^{(\varphi)}=- \frac{B_1}{4\pi} \, Y(r,\theta) \, r^{k-3} \,
\frac{\left(1-\frac{2 \, a
\,L}{r^3}\right)^k}{\left(1-\frac{2}{r}\right)^{k/2}} \, \sin^{k-1}
\theta
\end{eqnarray}
in which
\begin{eqnarray*}
Y(r,\theta)=\left\{(k-2)\cot^2\theta - 2
\left(1-\frac{3}{r}\right)+k \frac{\left(1-\frac{3}{r}\right)^2}
{\left(1-\frac{2}{r}\right)}\right.\nonumber\\
\left.+\frac{\frac{6 \, a \, L}{r^3}}{\left(1-\frac{2 \, a
\,L}{r^3}\right)}\left[-5+\frac{7}{r}+2\,k\left(1-\frac{3}{r}\right)\right]\right\}.\qquad
\end{eqnarray*}
We have derived two different expressions for the current density
$J^{(\varphi)}$ [equations (\ref{J_phi_Ohm}) and (\ref{J_phi_Max})].
Evidently, they have to be consistent
\begin{eqnarray}\label{consistency equ}
V^{(r)} A_1(r)+\left(1-\frac{2}{r}\right) \cot\theta\
\tilde{V}^{(\theta)}=\frac{1}{4 \pi \sigma u^0} \frac{Y
(r,\theta)}{r}.
\end{eqnarray}
Employing the assumption $V^{(r)}=C_1(\theta) \tilde{V}^{(\theta)}$
[equation (\ref{assum_Vr})] and the definition of $u^0$ [equation
(\ref{u0})], in the above equation, it gives the meridional velocity
as
\begin{eqnarray}\label{meridional velocity}
\tilde{V}^{(\theta)}=S_0 \sqrt{I}\ Y,
\end{eqnarray}
wherein $S_0=\frac{-1}{4\pi\sigma}$ has the dimension of magnetic diffusivity and
\begin{eqnarray*}
I&=&\frac{1-\left[V^{(\varphi)}\right]^2}{S_1^2+(S_0 \, Y)^2
\left[\left(1-\frac{2}{r}\right)+C_1^2(\theta)\right]},\\
S_1&=&A_1(r)\, r \left(1-\frac{2}{r}\right)^{-1/2} C_1(\theta)+r
\left(1-\frac{2}{r}\right)^{1/2} \cot\theta.
\end{eqnarray*}
Function $Y$ has a zero in a point between $k=2$ and $k=3$ for all
grid points in Figure \ref{fig_Disc_config}. For $k>2$, it is positive
($Y>0$) and for $k\leq 2$, it is negative ($Y<0$). If we choose the
second set ($k\leq 2$ and $Y<0$), then as above, $S_0$ must be
negative, so that the vertical velocity is positive to indicate
inflows. If the first set ($k>2$ and $Y>0$) is chosen, the only
difference will be in the sign of $S_0$ that must be positive.
Continuation of the story is the same as other set.

\begin{figure}
\begin{center}
\includegraphics[width=7cm,height=7cm]{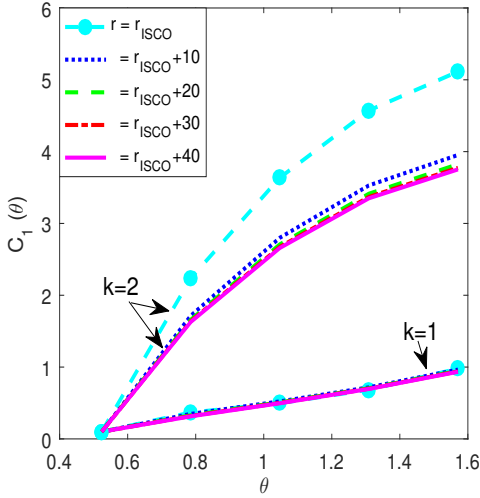}
\caption{Profiles of $C_1(\theta)$ at the different dimensionless
radial distances shown in the key for two values of $k$. The other
constant parameters are $a=0.1,\ l=1,\ \sigma=10,\
\dot{M}_0=-10^{-8} \frac{M_{\odot}}{year}$.}\label{fig_C1}
\end{center}
\end{figure}

Up to now, both the radial and meridional velocities and the mass
accretion rate have been obtained in terms of $C_1(\theta)$. To
define this unknown function, we aid from the integrability
condition of pressure
$$\frac{\partial^2 \ p}{\partial r \ \partial \theta}=\frac{\partial^2 \ p}{\partial \theta  \ \partial r}.$$
It provides a second-order ordinary differential equation for
$C_1(\theta)$ as
$$\frac{d^2 C_1(\theta)}{d\theta^2}=\Psi (r,\theta),$$
where $\Psi (r,\theta)$ is a known function of $r$ and $\theta$ in
terms of $C_1(\theta)$ and $\frac{d C_1(\theta)}{d\theta}$ that have
been derived in Appendix \ref{Appendix A}. The above differential
equation can be solved numerically with appropriate boundary
conditions. Integration is initiated from the upper boundary surface
(i.e. $\theta=\pi/6$) with the following boundary condition
\begin{eqnarray*}
C_1(\theta) \, |_{\theta=\frac{\pi}{6}}=0.1, \qquad \& \qquad
\frac{d C_1(\theta)}{d\theta}|_{\theta=\frac{\pi}{6}}=1.
\end{eqnarray*}
After running a complicated code, $C_1$ is achieved as an ascending
function of $\theta$ (Figure \ref{fig_C1}). $C_1$ has been presumed to
be dependent just on $\theta$. Thus, we have to choose the set of
free parameters as well as the boundary conditions in the manner
that $C_1$'s profiles in different radii are coincident. Otherwise,
the relevant radii must omit from our allowed radial interval. This
point determines for us the allowed radius for the inner edge of our
disc. For instance, we discuss two sets of free parameters employed
in plotting Figure \ref{fig_C1}. For the set $k=1$, we choose
$r_{ISCO}$ as the inner edge (i.e. $r_{in}=r_{ISCO}$). However, for
the set $k=2$, we choose $r_{in}=r_{ISCO}+10$. On account of this
fact that for $k=2$, $C_1$'s profile in $r_{ISCO}$ is not coincident
on the others, but 10 units farther than ISCO, our expectation about
independency of $C_1$ on the radial distance $r$ is almost satisfied.

With specified $C_1(\theta)$, we can obtain the radial and
meridional velocities, and mass accretion rate through equations
(\ref{assum_Vr}), (\ref{meridional velocity}), and (\ref{mass
accretion2}), respectively. Both radial velocity and mass accretion
rate must be negative. This negativity indicates the inflow towards
the central black hole. Because, the positive radial direction is
defined in the direction of increasing $r$. Moreover, the positive
meridional direction is defined in the direction of increasing
$\theta$ too (Figure \ref{fig_Disc_config}). Therefore, the negative
meridional velocity denotes outflow which is beyond the scope of
this paper. Whereas the radial inflow velocity must be negative and
the meridional velocity must be positive, then we rewrite the
equation (\ref{assum_Vr}) as
\begin{eqnarray*}
V^{(r)} = - C_1(\theta) \, \tilde{V}^{(\theta)}.
\end{eqnarray*}

At this time, there remains just two unknown physical variables, gas
pressure and total density. Gas pressure may be achieved from the
pressure gradient terms in momentum equations [(\ref{Radial equ1}) or
(\ref{Meridional equ1})]. After some manipulations, these two
equations change to equations (\ref{Radial_pressure_gradient}) and
(\ref{Meridional_pressure_gradient}). We prefer to employ the radial
component of pressure gradient [equation
(\ref{Radial_pressure_gradient})], due to its simplicity in
integration. Because, $C_1(\theta)$ behaves like a constant in
radial integration. It can be rewritten as
\begin{eqnarray}\label{Radial_pressure_gradient2}
\frac{\partial p}{\partial r}= \chi (r,\theta).
\end{eqnarray}
In fact, we rename the right side of equation
(\ref{Radial_pressure_gradient}) as $\chi (r,\theta)$, which is a
known function of $r$ and $\theta$. Now, the gas pressure as a
function of $r$ and $\theta$ is obtained by integrating equation
(\ref{Radial_pressure_gradient2}) with respect to the radial
distance,
\begin{eqnarray}
p(r,\theta)=p_{0_{mag}}^S+\int_{r_{0_{mag}} \, = \, r_{ISCO}-1}^r \,
\chi(r,\theta)\, dr.
\end{eqnarray}
Within the magnetosphere surrounding the black hole, the magnetic
pressure dominates. We benefit from this fact to define the
integration constant $p_{0_{mag}}^S$ as the magnetic pressure of the
radius $r_{0_{mag}}$ in the magnetosphere. Besides, total density can
be attained through the definition of mass accretion rate [equation
(\ref{mass accretion1})]
\begin{eqnarray}\label{rho}
\rho (r,\theta)=\frac{\dot{M}}{r^2 \sin\theta \, V^{(r)}}
\frac{\left(1-\frac{2 a
L}{r^3}\right)}{\left(1-\frac{2}{r}\right)}\left(1-V^2\right)-p.
\end{eqnarray}

At present, all physical functions of the disc have been specified
in terms of some free parameters (i.e. $\sigma$, $l$, $a$,
$\dot{M}_0$ and $k$). An appropriate set of these free parameters is
a set that both $C_1$ does not vary significantly with the radial
distance $r$ and the total density is positive throughout the disc.
During running the code, it is possible to encounter the situations
that for a specific set of free parameters, $C_1$ is not necessarily
independent on $r$ (Figure \ref{fig_C1}) and or the density is
negative from a certain radius to the next. Obviously, these
features are undesirable to have physically meaningful disc
solutions. Hence, in order to avoid these unpleasant cases, we
forced to limit the allowed interval for the disc's radius.

In this interval, both $C_1$ has to independent on $r$ and total density
must be positive. It means that if the profile of $C_1$ in a special
radius does not coincident on that of $C_1$ in other radii and or total density is
negative there, then that special radius must omit from our allowed
radial interval. For the set $k=1$, the inner edge of the disc rests
on ISCO. However, for the other set ($k=2$), disc starts off 10
units farther than ISCO. Because next to ISCO, $C_1$'s profile seems
to be dependent upon $r$ (Figure \ref{fig_C1}). Besides the inner
edge, these two sets of solutions about the outer edge have
significant discrepancy too. Occasionally, after a specific radius,
it is possible that the total density becomes negative. It results in
truncating the disc there. This is the event happened in the case of
$k=1$, and restricts the disc's outer radius. That is why the disc
in this case is radially shorter than the other case (Figure
\ref{fig6}a$_1$ and b$_1$).

\begin{figure}
\begin{center}
\includegraphics[width=9cm,height=12cm]{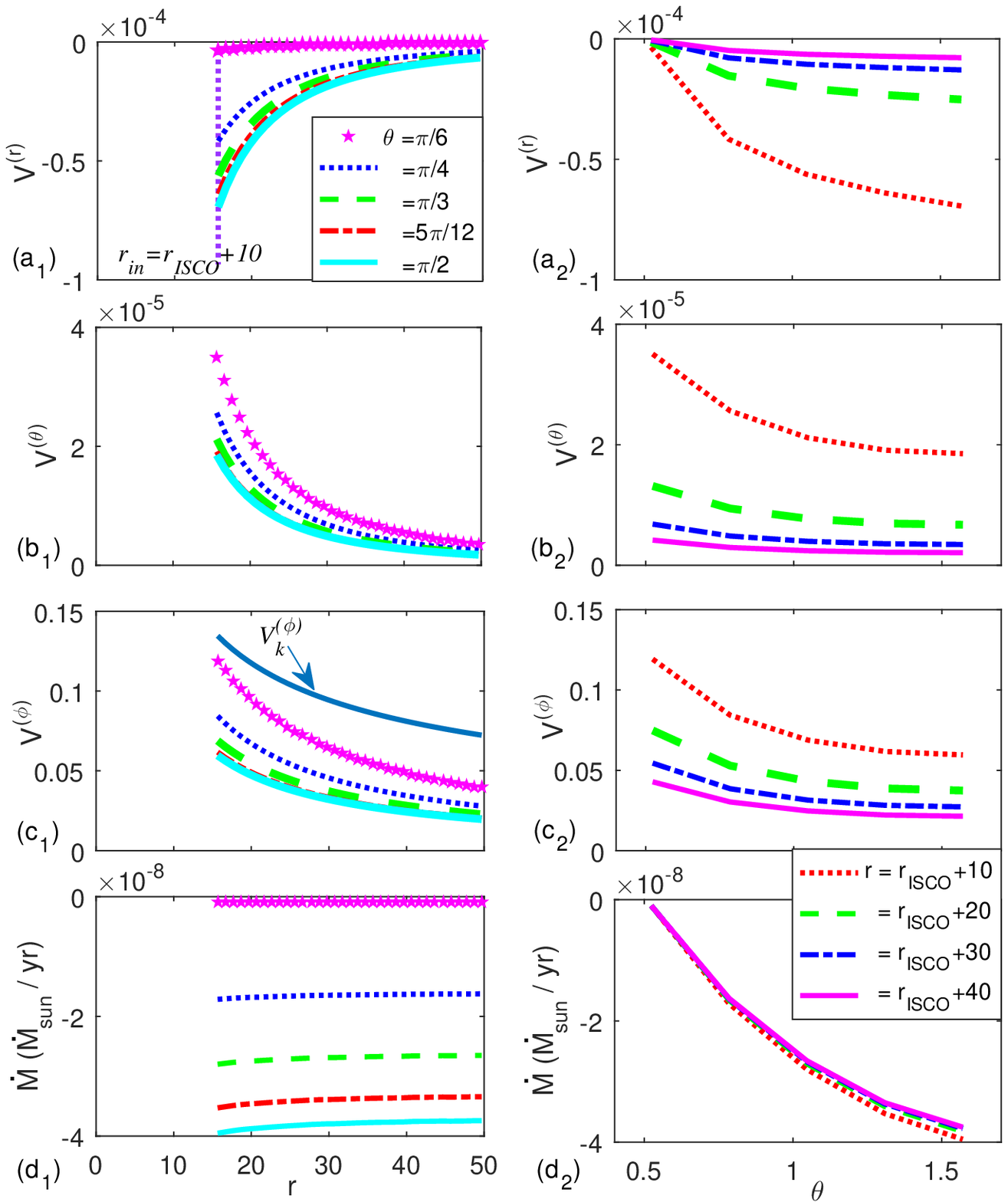}
\includegraphics[width=9cm,height=6cm]{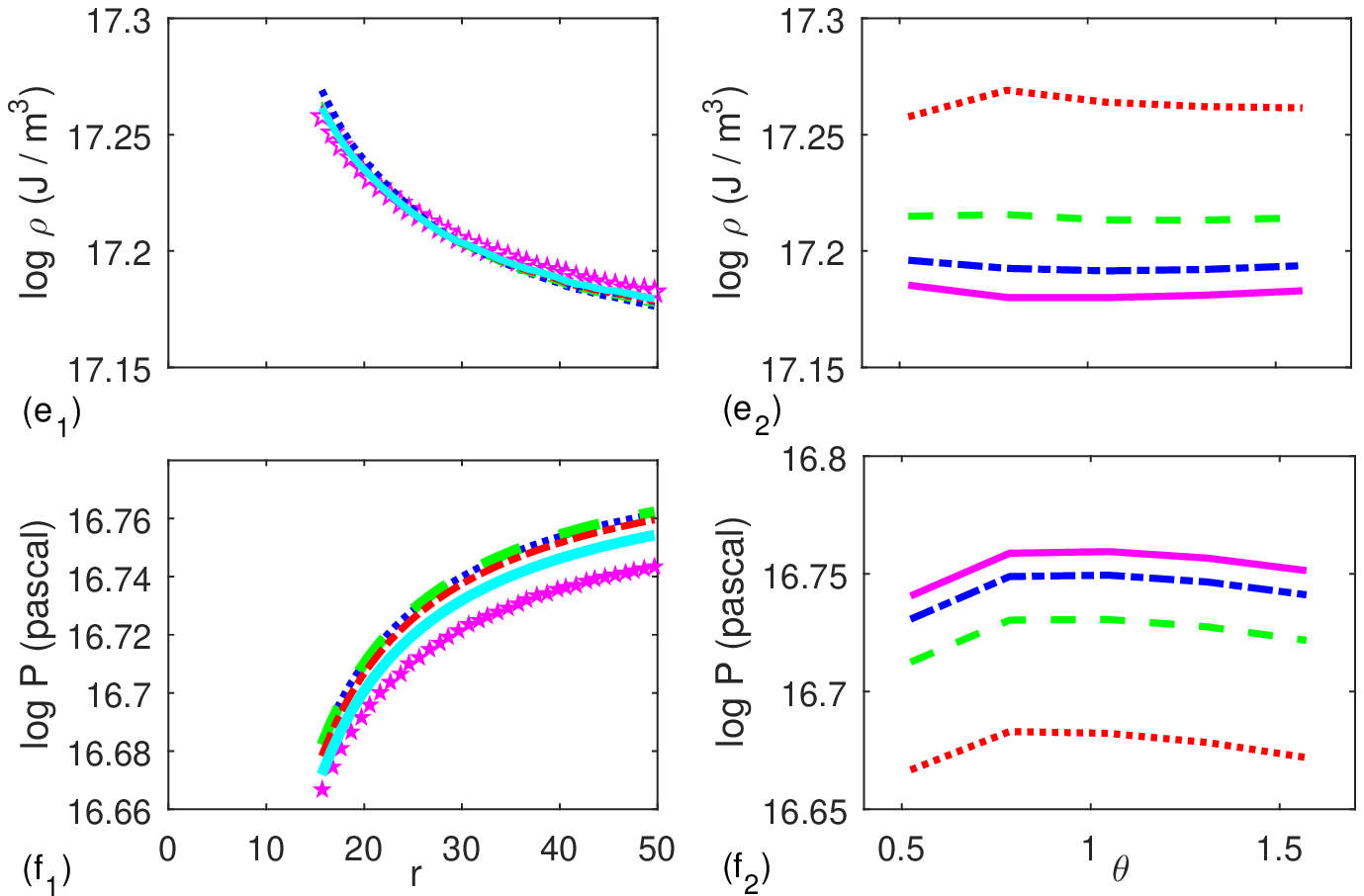}
\caption{Variation of (a) radial, (b) meridional and (c) azimuthal
velocities, in addition to (d) mass accretion rate, (e) total density
and (f) gas pressure along the radial direction $r$ at the different
meridional layers shown in the key (left column) and also along the
meridional direction $\theta$ at the different radii shown in the
key (right column). The constant parameters are $a=0.1$, $l=1$,
$\sigma=10$, $k=2$ and $\dot{M}_0=-10^{-8}
\frac{M_{\odot}}{year}$.}\label{fig5}
\end{center}
\end{figure}

\begin{figure}
\begin{center}
\includegraphics[width=8.5cm,height=6cm]{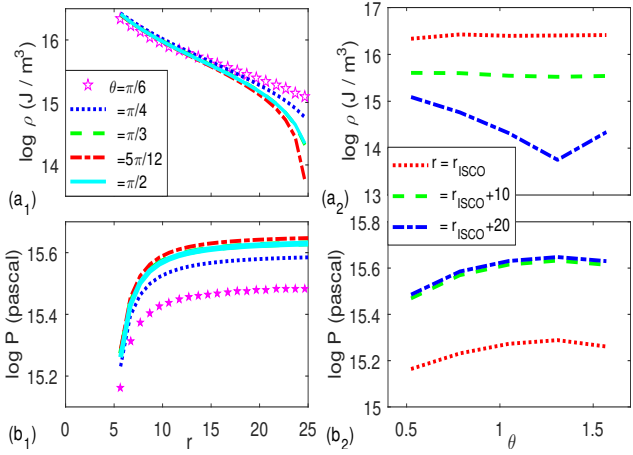}
\caption{Repetition of the last two rows of Fig. \ref{fig5}, but
for $k=1$.}\label{fig6}
\end{center}
\end{figure}

Figure \ref{fig5} gives both radial (left column) and meridional
(right column) variations of all the physical quantities of the
disc. Close to the inner edge of the disc, all three components of
fluid velocity are extremely high and gradually fall off radially
outwards (Figure \ref{fig5}a$_1$, b$_1$, and c$_1$). Azimuthal
velocity of the surface layer ($\theta=\frac{\pi}{6}$) near ISCO is
nearly super-Keplerian and reaches to sub-Keplerian regime in the
inner edge ($r_{in}=r_{ISCO}+10$). Thus, in our allowed radial
interval for $k=2$ solution set, rotation of the disc is
sub-Keplerian all over the disc (Figure \ref{fig5}c$_1$). The
horizontal constant lines in Figure \ref{fig5}d$_1$ are a firm
confirmation on the radial independency of mass accretion rate. As
we get radially closer to the central black hole, the disc becomes
denser (Figure \ref{fig5}e$_1$); however, its gas pressure falls off
rapidly (Figure \ref{fig5}f$_1$).

From the disc surface ($\theta=\frac{\pi}{6}$) towards the equator
($\theta=\frac{\pi}{2}$), radial inflow including radial velocity
(Figure \ref{fig5}a$_2$) and mass accretion rate (Figure \ref{fig5}d$_2$) becomes faster.
Nonetheless, the meridional (Figure \ref{fig5}b$_2$)
and azimuthal (Figure \ref{fig5}c$_2$) velocities slow down.
The surface layer has a super-Keplerian rotation near ISCO and
sub-Keplerian one in the outer edge. The other layers obey the
sub-Keplerian regime all over the allowed radius. While the total density
remains nearly constant along the meridional direction (Figure
\ref{fig5}e$_2$), pressure ascends from the surface up to around
$\theta=\frac{\pi}{4}$, then it becomes constant (Figure \ref{fig5}f$_2$).

For the solution set $k=1$, as mentioned above, disc starts on ISCO
up to around $r=25$. It means that the radius of the disc shrinks in
this case respect to the other set (Figure \ref{fig6}a$_1$ and
b$_1$). Mass accretion rate and fluid velocity behave in the same manner of the
solution set of $k=2$ in the radial and meridional directions.
Comparing Figure \ref{fig6}a$_1$ with Figure \ref{fig5}e$_1$, it is seen that
the ascending behaviour of the total density in the radial direction for the $k=1$ solution set
seems to be different with another solution set. Meridional behaviour of the total density in
inner region is constant like the $k=2$ solution set. However, in outer region ($r=r_{ISCO}+20$),
the total density finds the meridional angular dependency (Figure \ref{fig6}a$_2$).
In both solution sets, pressure is an ascending function of the
radial distance. For the set $k=1$, as $r$ increases, pressure tends
to remain constant in outer region after an initial ascent (Figure \ref{fig6}b$_1$). However, for the other set ($k=2$),
pressure ascends rapidly towards the outer edge (Figure \ref{fig5}f$_1$). The meridional behaviour of pressure is just a little
different for both sets. It rises uniformly from the surface towards
the equator (Figure \ref{fig6}b$_2$).

\begin{figure}
\begin{center}
\includegraphics[width=9cm,height=9cm]{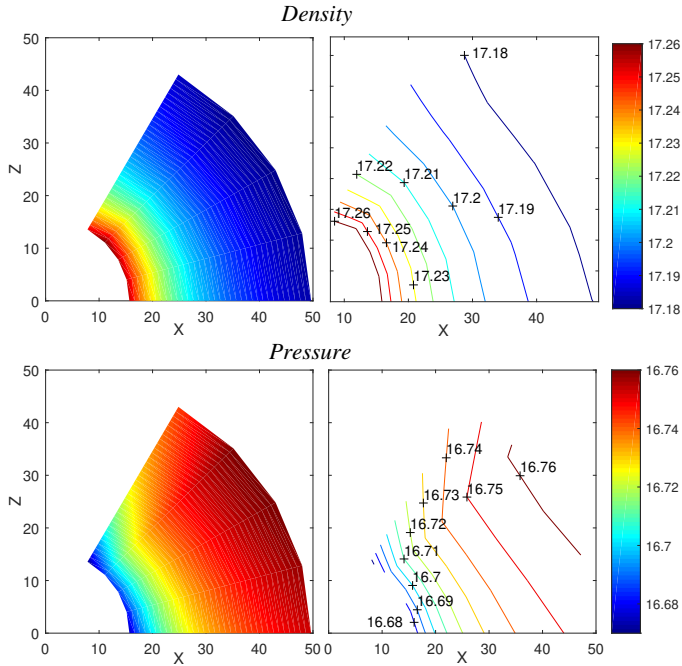}
\caption{Density and pressure colored distributions and contours in
meridional plane. The constant parameters are the same as Fig.
\ref{fig5}.}\label{fig_Contours}
\end{center}
\end{figure}

Density and pressure coloured distributions has been plotted in Figure \ref{fig_Contours},
as a strong verification on interpretations of
profiles of density and pressure in Figure \ref{fig5}e$_1$, e$_2$,
f$_1$, and f$_2$.

\begin{figure}
\begin{center}
\includegraphics[width=9cm,height=9cm]{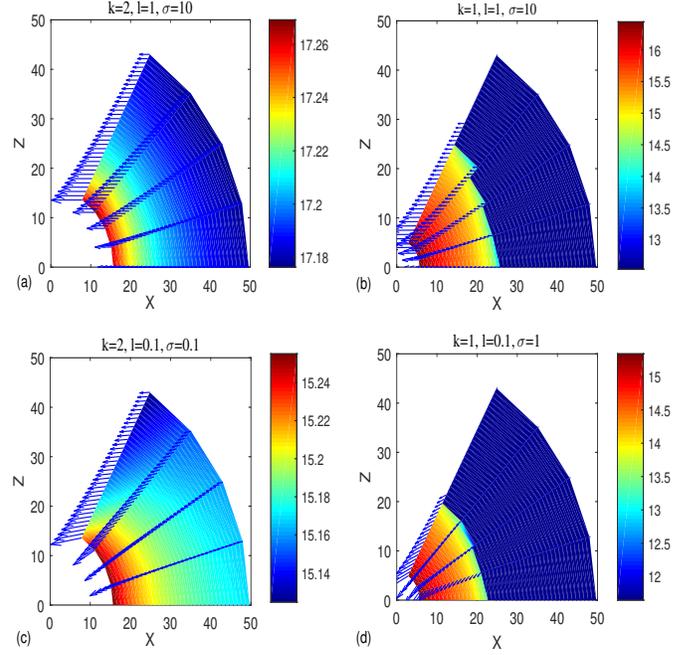}
\caption{Density colored distributions and meridional flow pattern
for different values of $k$, $l$ and $\sigma$ written in the title
of each panel.}\label{fig_vector_field}
\end{center}
\end{figure}

In Figure \ref{fig_vector_field}, fluid flow has been represented in
meridional plane by arrows. The length and direction of the vectors
indicate the magnitude and orientation of the total fluid velocity, respectively.
Density coloured distribution is seen in the background of this
figure as well. The dark blue colour in the right column panels
indicates the region with negative density that is a forbidden
region. It demonstrates that for the $k=1$ solution set (Figure
\ref{fig_vector_field}b and d), the radius of the disc shrinks
with respect to the other set (Figure \ref{fig_vector_field}a and c) and
the outer edge becomes nearer to the inner edge resting on ISCO.
When rotation of the disc (Figure \ref{fig5}c$_1$) is a few orders of
magnitude faster than the inflow velocity (Figure \ref{fig5}a$_1$ and
b$_1$), plasma flows in the azimuthal direction (Figure
\ref{fig_vector_field}a and b). It likes an equilibrium toroidal
configuration around the central black hole. As $l$ and $\sigma$
decrease, azimuthal and inflow velocities become comparable. This is
quite obvious from the vectors' direction (Figure
\ref{fig_vector_field}c and d).

\begin{table*}
 \centering
 \begin{minipage}{140mm}
  \caption{Physical constants and conversion factors between code and SI units.}
  \begin{tabular}{@{}llll@{}}
  \hline
  \hline
   Constant in SI units &  & & Code value\\
 \hline
 $G=6.67384\times 10^{-11}\ N.m^2/kg^2$ & & & $G=1$\\
 $M=10 M_{\odot}=20\times 10^{30}\ kg$ & & & $M=1$\\
 $c=3\times 10^8\ m/s$ & & & $c=1$\\
 $\mu=1260\times10^{17}\ T.m^3$ & & & $\mu=5.2\times 10^{-11}$\\
 \hline\hline
Quantity  &  SI units   &   Geometric units   &   Conversion factor to SI units\\
\hline
Length & $m$ & $G M c^{-2}$ & $1.4831\times10^4 \ \ m$\\
Time & $s$ & $G M c^{-3}$ & $4.9436\times10^{-5} \ \, s$\\
Velocity & $m/s$ & $c$ & $3\times 10^8 \ \ m/s$\\
Accretion rate & $kg/s$ & $c^3 G^{-1}$ & $4.0456\times 10^{35} \ \ kg/s$\\
Rest mass density& $kg/m^3$ & $c^6 G^{-3} M^{-2}$ & $6.1311\times 10^{18} \ \ kg/m^3$\\
Internal energy per unit volume& $J/m^3$ & $c^8 G^{-3} M^{-2}$ & $5.5180\times 10^{35} \ \ kg/(m.s^2)$\\
Pressure& $pascal$ & $c^8 G^{-3} M^{-2}$ & $5.5180\times 10^{35} \ \ kg/(m.s^2)$\\
Electrical conductivity& $1/s$ & $c^3 G^{-1} M^{-1}$ & $2.0228\times10^4 \ \ \, 1/s$\\
Magnetic field& $T$ & $c^4 G^{-3/2} M^{-1}$ & $7.4283\times10^{17} \ \ T$\\
Magnetic dipole moment& $T.m^3$ & $G^{\, 3/2} M^2 c^{-2}$ & $2.4232\times 10^{30} \ \ T.m^3$\\
\hline
\hline
\end{tabular}
\end{minipage}
\end{table*}

\subsubsection{Effect of free parameters}
It is time to discuss the properties and the physical implications
that our achieved solutions involve. To conceive the role of free
parameters on MHD behaviour of the disc, we plot the
meridional dependency of the physical functions with respect to
different values of these parameters. Incidentally, to have a better
physical sense and direct interpretation, we plot them in physical
units (SI), with the help of conversion factors calculated in Table
1.

\begin{figure}
\begin{center}
\includegraphics[width=8cm,height=9cm]{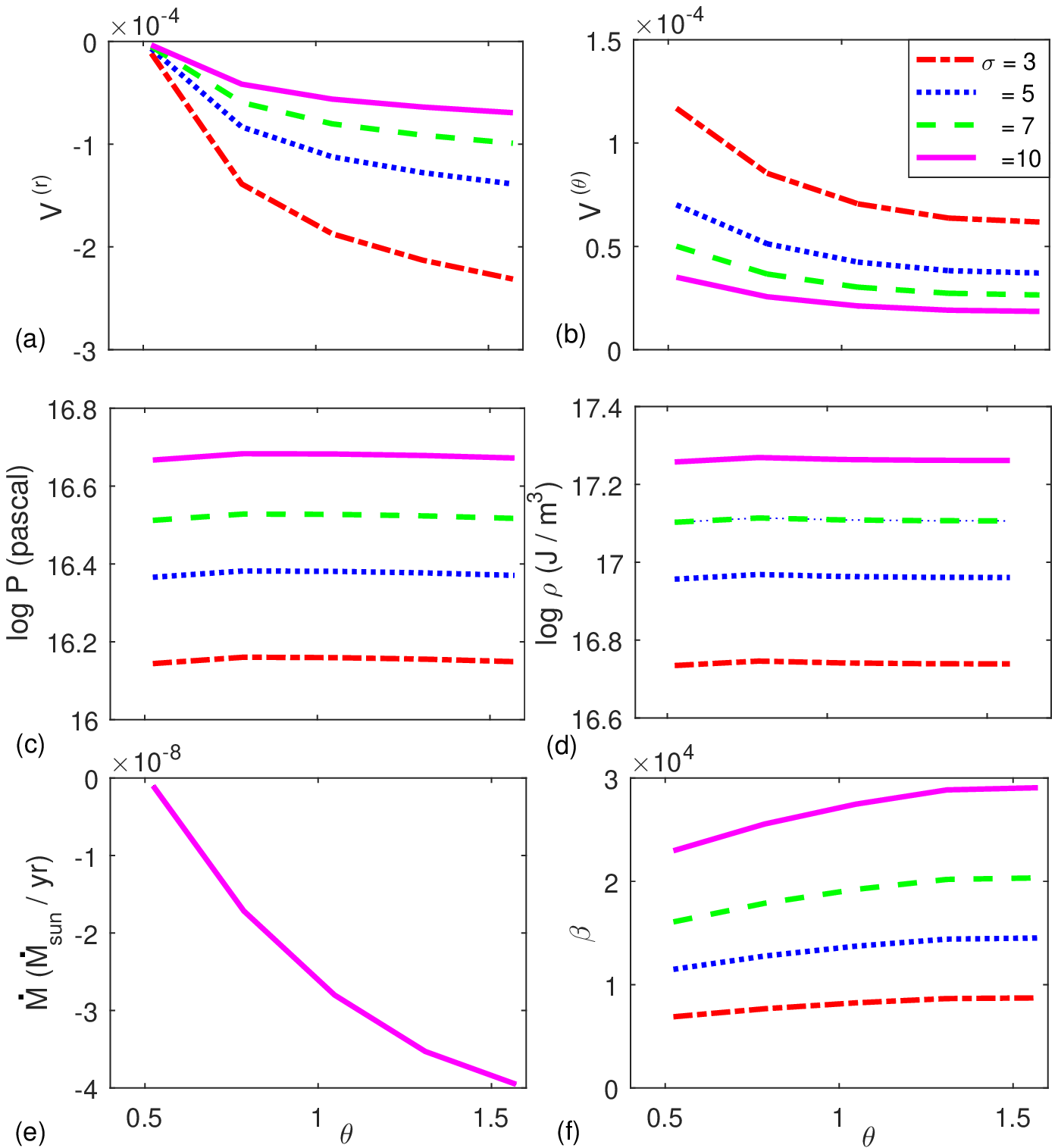}
\caption{Meridional variation of (a) radial
and (b) meridional velocities, (c) pressure (d) total density (e) mass
accretion rate and (f) ratio of the gas to magnetic pressure
$\beta$, at radius $r_{in}$ for different values of $\sigma$ shown
in the key. The other constant parameters are $a=0.1$, $l=1$, $k=2$
and $\dot{M}_0=-10^{-8} \frac{M_{\odot}}{year}$.}\label{fig_sigma}
\end{center}
\end{figure}

Effect of electrical conductivity on some impressible physical
variables has been represented in Figure \ref{fig_sigma}. Once
conductivity grows large or resistivity becomes small, radial (Figure
\ref{fig_sigma}a) and meridional (Figure \ref{fig_sigma}b) fluid
velocities slow down. While, gas pressure (Figure \ref{fig_sigma}c)
and total density (Figure \ref{fig_sigma}d) exceed, mass accretion rate
(Figure \ref{fig_sigma}e), rotational velocity, and magnetic pressure
are not affected by resistivity at all. Magnetic pressure
invariability and gas pressure ascent as $\sigma$ goes up result
in raising the ratio of gas to magnetic pressure $\beta$ (Figure
\ref{fig_sigma}f).

\begin{figure}
\begin{center}
\includegraphics[width=8cm,height=6cm]{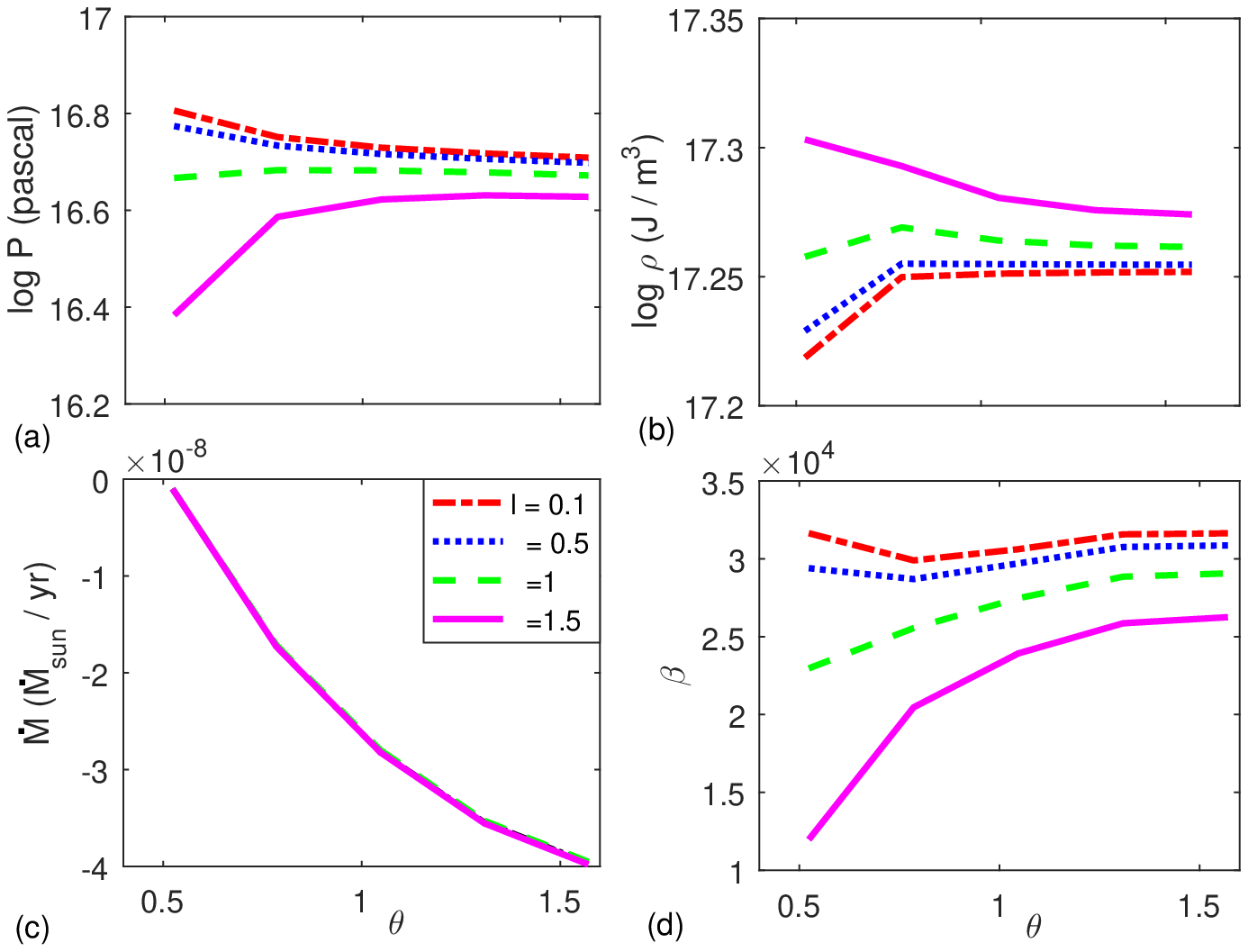}
\caption{Meridional variation of (a) pressure
(b) total density (c) mass accretion rate and (d) ratio of the gas to magnetic pressure $\beta$,
at radius $r_{in}$ for different values of $l$
shown in the key. The other constant parameters are $a=0.1$,
$\sigma=10$, $k=2$ and $\dot{M}_0=-10^{-8}
\frac{M_{\odot}}{year}$.}\label{fig_l}
\end{center}
\end{figure}

Disc's rotation just influences the gas pressure, total density, and
subsequently $\beta$ (Figure \ref{fig_l}). In other words, radial and
meridional velocities as well as mass accretion rate (Figure
\ref{fig_l}c) and magnetic pressure are invariant relative to
angular momentum parameter $l$. As disc rotates faster, gas pressure
decreases (Figure \ref{fig_l}a) and evidently via the equation
(\ref{rho}), disc becomes denser (Figure \ref{fig_l}b). Obviously, it
results in falling off $\beta$ in this occasion (Figure \ref{fig_l}d).

\begin{figure}
\begin{center}
\includegraphics[width=8cm,height=9cm]{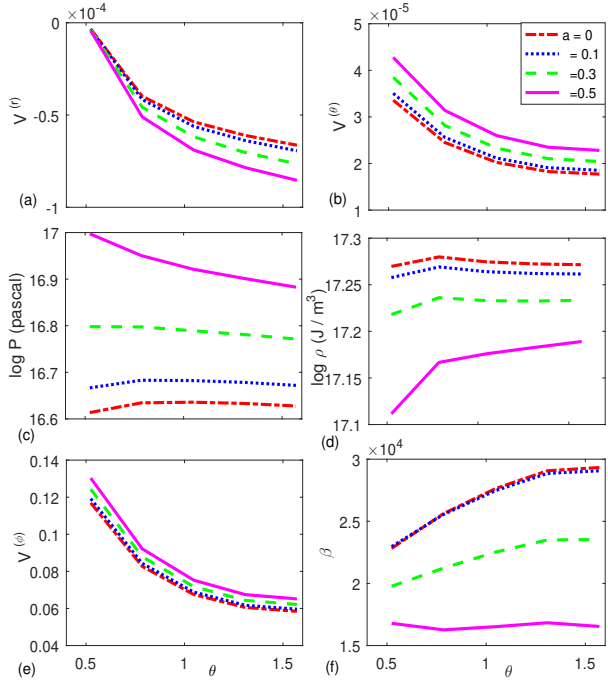}
\caption{Meridional variation of (a) radial and
(b) meridional velocities, (c) pressure (d) total density (e) azimuthal
velocity and (f) ratio of the gas to magnetic pressure $\beta$, at
radius $r_{in}$ for different values of $a$ shown in the key. The
other constant parameters are $l=1$, $k=2$, $\sigma=10$ and
$\dot{M}_0=-10^{-8} \frac{M_{\odot}}{year}$.}\label{fig_a}
\end{center}
\end{figure}

Although inflow velocity including radial and meridional components
of fluid velocity is not impressed by rotation of the disc, they
are affected by the spin of the central black hole. As central
object spins faster, inflow velocity becomes faster (Figure
\ref{fig_a}a and b) and disc rotates faster too (Figure
\ref{fig_a}e). Gas pressure heightens (Figure \ref{fig_a}c), while
density falls off (Figure \ref{fig_a}d). Descending behaviour of $\beta$
(Fig. \ref{fig_a}f) indicates that the increase in magnetic pressure
with accelerating the spin of the black hole is more than gas
pressure.

\begin{figure}
\begin{center}
\includegraphics[width=8cm,height=6cm]{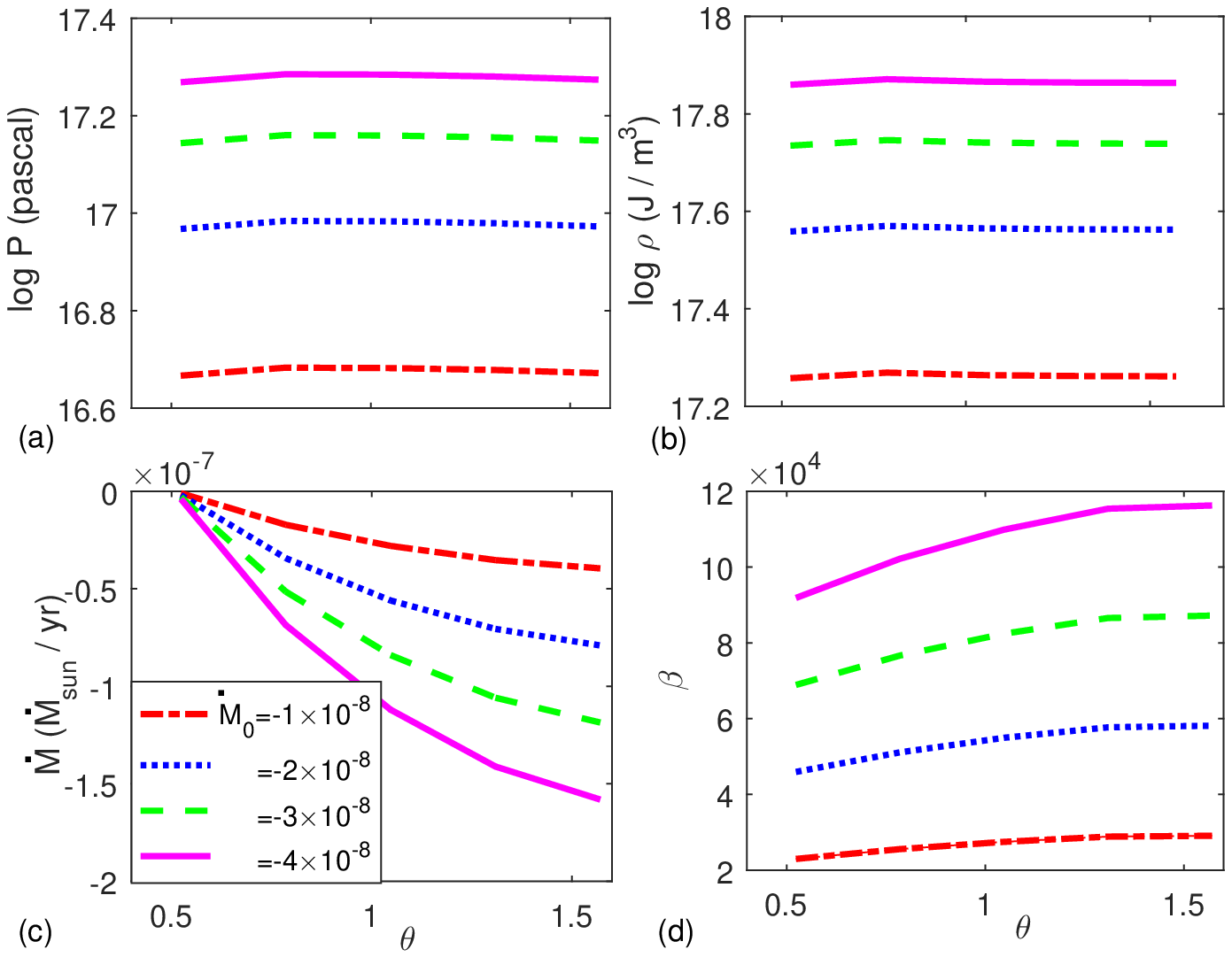}
\caption{Meridional variation of (a)
pressure (b) total density (c) mass accretion rate and (d) ratio of the gas to magnetic pressure $\beta$,
at radius $r_{in}$ for different
values of $\dot{M}_0$ shown in the key. The other constant
parameters are $a=0.1$, $l=1$, $\sigma=10$, $k=2$.}\label{fig_Mdot}
\end{center}
\end{figure}

Larger the value of mass accretion rate constant $\dot{M}_0$ (Figure
\ref{fig_Mdot}c), higher are the values of gas pressure (Figure
\ref{fig_Mdot}a), total density (Figure \ref{fig_Mdot}b), and the
ratio of gas to magnetic pressure $\beta$ (Figure \ref{fig_Mdot}d).

\begin{figure}
\begin{center}
\includegraphics[width=8cm,height=9cm]{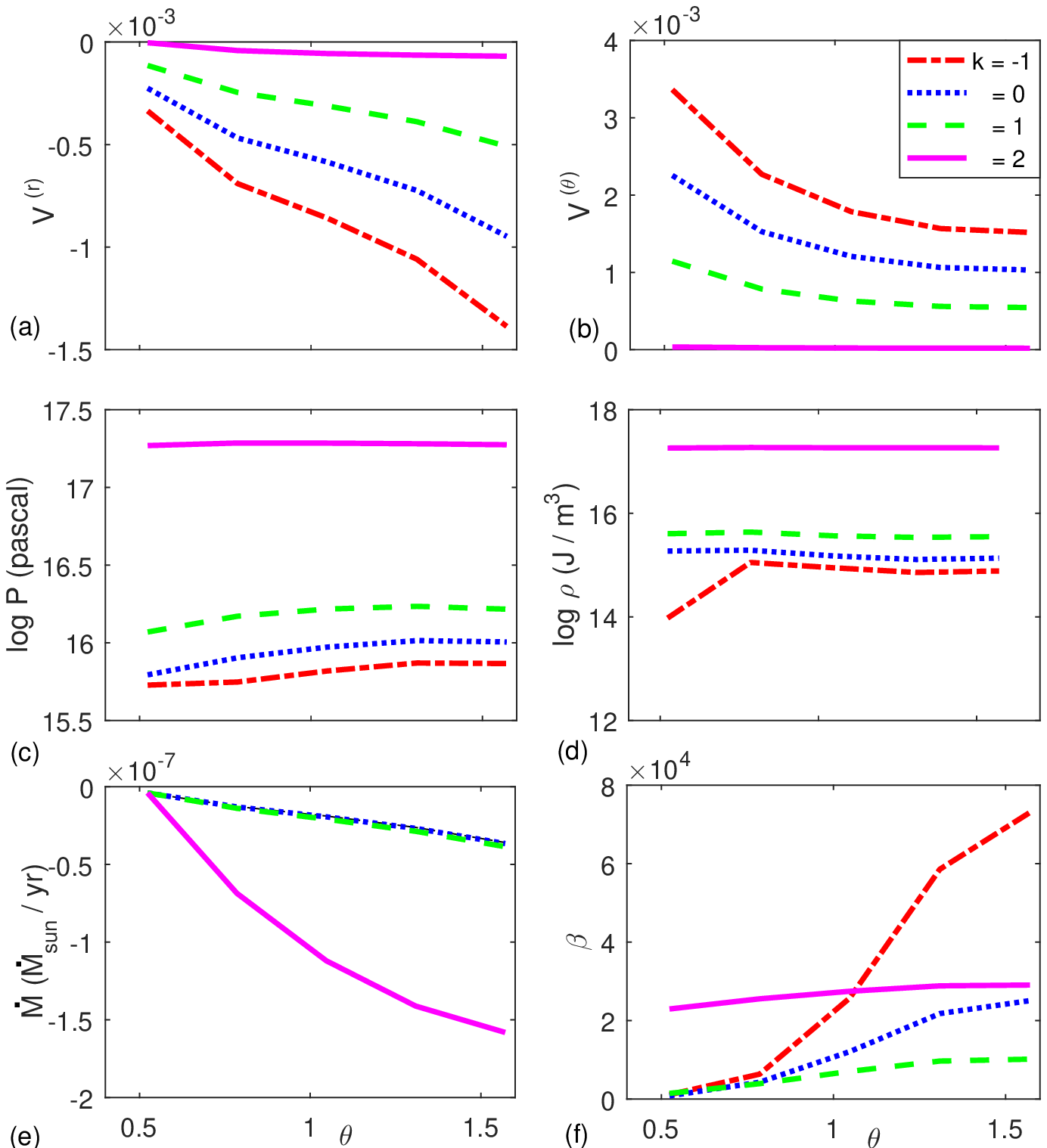}
\caption{Meridional variation of (a) radial and
(b) meridional velocities, (c) pressure (d) total density (e) mass
accretion rate and (f) ratio of the gas to magnetic pressure
$\beta$, at radius $r_{in}$ for different values of $k$ shown in the
key. The other constant parameters are $l=1$, $a=0.1$, $\sigma=10$ and
$\dot{M}_0=-10^{-8} \frac{M_{\odot}}{year}$.}\label{fig_k}
\end{center}
\end{figure}

$k$ ascent leads to decelerate the radial (Figure \ref{fig_k}a) and
vertical (Figure \ref{fig_k}b) inflow velocities and heightening the
gas pressure (Figure \ref{fig_k}c), total density (Figure \ref{fig_k}d), and
mass accretion rate (Figure \ref{fig_k}e). In addition, it results in
sharp variations in $\beta$ around the equator (Figure \ref{fig_k}f).

\section{Discussion and conclusion}
Analytical and numerical (semi-analytical) investigations of thick accretion discs
around a rotating compact object in presence of an external dipolar
magnetic field considering all three components of the fluid velocity
have not been carried out in any detail so far. In
this paper, we have developed an axisymmetric stationary
two-dimensional model of the magnetised tori accreted from the
resistive plasma surrounding a rotating black hole. Importance of
the general relativity in the discussions of accretion physics
around a black hole is no secret to anyone and the full relativistic
treatment is required. Consequently, we have derived the governing
MHD equations in the full general relativistic
framework. They are a coupled set of highly non-linear equations
that are in general so difficult to solve. To avoid the mathematical
complexities, we employ the linear order approximation on the Kerr
parameter $a$ and ignore the effects of radiation field.

These simplifying assumptions are indeed considered as practical
approximations applicable to the full general relativistic MHD system.
Additionally, assumption of similar radial dependency for the radial and
meridional velocities resulting in a simple form for mass accretion
rate helps a lot in solving the equations.

Moreover, considering a special model for the toroidal magnetic field leads to the fact that the
radial and meridional components of the 4-vector current density are
vanished. The azimuthal current produced owing to the motion of the
magnetofluid generates a poloidal magnetic field inside the
disc as well. It has been depicted that the disc's poloidal magnetic field
and the spin of the central black hole are strongly related.
Connection of the disc field to the external dipolar field occurs
due to the presence of the finite conductivity for the plasma. The
meridional structure of the disc is mainly on account of the balance
of plasma pressure gradient, magnetic force due to the poloidal
magnetic field, and the centrifugal force. In our scenario, different
free parameters are determinant and play a crucial role in
calculating the accretion tori features and have a major influence
on process of accretion around a black hole.

In conclusion, we find that the self-consistent equilibrium
solutions in the relativistic framework do exist for a slowly
rotating black hole with a dipolar magnetic field accreting matter
from a disc having all the components of velocity non-zero without
invoking any thin disc approximations. The existence of such an
equilibrium structure encourages one to put aside our simplifying
assumptions gradually and look for generalisations of the analysis
to a rapidly rotating central black hole. On the other hand, yet
another important aspect to be considered is the generalisation of
the analysis to cases where a toroidal magnetic field is generated
by the inertia of the plasma and backward bending the external
dipolar magnetic field lines. This toroidal magnetic field is
associated with a hoop stress that can collimate a hydromagnetic
outflow over large distances and form a jet.

This work can also be useful for the general relativistic
MHD simulations that suffer mainly from the lack of
exact analytical and semi-analytical solutions to use as initial conditions and to
compare their achieved results.

\begin{appendix}
\section {Magnetic Field Lines} \label{Appendix B}
Magnetic field lines are curved lines drawn in space in such a way
that a tangent line at any point is parallel to the magnetic field
vector at that point. Slop of this line in any point ($r$,
$\theta$), in spherical coordinate, is defined by
$\frac{dr}{r\,d\theta}$. On the other hand, slop of the magnetic
field in that point is defined by $\frac{B_{(r)}}{B_{(\theta)}}$.
Thus, we have
\begin{eqnarray}\label{Br/Bt}
\frac{dr}{r\,d\theta}=\frac{B_{(r)}}{B_{(\theta)}}.
\end{eqnarray}
It gives
\begin{eqnarray}\label{Br/Bt_BH}
-2 \cot\theta \, d\theta=\frac{dr}{r},
\end{eqnarray}
for poloidal magnetic filed of the central black hole in the limit
($\frac{m}{r}<<1$), and gives
\begin{eqnarray}\label{Br/Bt_Disc}
-\cot\theta \, d\theta=\left[\frac{1}{r}+\frac{\frac{6\, a
L}{r^4}}{1-\frac{2\, a L}{r^3}}\right] dr,
\end{eqnarray}
for the disc. On account of the linearised approximation on
$a$, it is good to notify that the terms including $a^2$
have been ignored. Integrating the differential equation (\ref{Br/Bt_BH}), it
yields
$$r=\frac{r_0}{sin^2\theta},$$
for the seed field. Do the same work for equation
(\ref{Br/Bt_Disc}), it provides an algebraic equation
$$r^2\left(r-\frac{r_0}{\sin\theta}\right)=2 \, a \, L.$$
Solving it, we have
$$r=\frac{1}{6}\,Q+\frac{2}{3}\,\frac{d^2}{Q}+\frac{1}{3}\,d,$$
for the disc field. Where
\begin{eqnarray}
&& Q=\left(108\, b+8\, d^3+12\, \sqrt{12\, d^3 b+81\,
b^2}\right)^{1/3},\\[.1cm]
&& d=\frac{r_0}{\sin\theta},\\[.1cm]
&& b=2\, a\, L,
\end{eqnarray}
and $r_0$ is a constant of integration. Considering the toroidal magnetic field and knowing the relation
$$\frac{dr}{B_{(r)}}=\frac{r d\theta}{B_{(\theta)}}=\frac{r \sin \theta d\varphi}{B_{(\varphi)}}$$
between the components of the magnetic field, we achieve
$$\varphi=\varphi_0+\frac{b_{\varphi}}{B_1}\frac{1}{k-1}\frac{r^{1-k}}{\sin^k\theta \cos\theta}.$$
To visualize the field line
structure, it is usual to transform to a Cartesian frame through the
relations $X=r\sin\theta\cos\varphi$, $Y=r\sin\theta\sin\varphi$ and $Z=r \cos\theta$.

\onecolumn
\section {Derivation of $\Psi (r,\theta)$} \label{Appendix A}
At first, rewrite the equation (\ref{mass accretion1}) as
$$\left(\rho+p\right)\left(1-V^2\right)^{-1}=\frac{\dot{M}_0 C_1(\theta)}{r^2 \sin\theta \, V^{(r)}\left(1-\frac{2}{r}\right)
 \left(1-\frac{2 \, a \, L}{r^3}\right)^{-1}},$$
and substitute it in momentum equations (\ref{Radial equ1}) and
(\ref{Meridional equ1}), to achieve the components of pressure
gradient
\begin{eqnarray}\label{Radial_pressure_gradient}
\frac{\partial p}{\partial
r}=\frac{-\dot{M}_0}{\left(1-\frac{2}{r}\right) \left(1-\frac{2 \, a
\,L}{r^3}\right)^{-1} r^2 \sin\theta} N_1(r,\theta)
+\left(1-\frac{2}{r}\right)^{-1/2} J^{(\varphi)} \left[B_{(\theta)}\left\{1-[V^{(\varphi)}]^2\right\}+B_{(\varphi)} V^{(\theta)} V^{(\varphi)} \right],
\end{eqnarray}
\begin{eqnarray}\label{Meridional_pressure_gradient}
\frac{\partial p}{\partial \theta}&=&\frac{-\dot{M}_0 \,
C_1(\theta)}{\sin\theta}\left\{\frac{1}{r} \left(1-\frac{2 \, a
\,L}{r^3}\right) N_3(r,\theta)+\frac{6 \,a\,
L}{r^5}\tilde{V}^{(\theta)}-\frac{L^2}{r^4 \sin^2\theta}
\left(1-\frac{2 \, a
\,L}{r^3}\right)^{-1}\frac{\cot\theta}{C_1(\theta)\tilde{V}^{(\theta)}}\right\}\nonumber\\
&-&r J^{(\varphi)}\left[B_{(r)}\left\{1-[V^{(\varphi)}]^2\right\}+B_{(\varphi)} V^{(r)} V^{(\varphi)}\right].
\end{eqnarray}
Then, calculate the pressure second derivatives
\begin{eqnarray}\label{d2pdthetadr}
&&\frac{\partial^2 \ p}{\partial \theta \
\partial r}=\frac{-\dot{M}_0}{\left(1-\frac{2}{r}\right)
\left(1-\frac{2 \, a \, L}{r^3}\right)^{-1} r^2
\sin\theta}\left\{-\cot\theta N_1(r,\theta)+ \frac{\partial
N_1(r,\theta)}{\partial \theta}\right\}+N_2(r,\theta),
\end{eqnarray}
\begin{eqnarray}\label{d2pdrdtheta}
&&\frac{\partial^2 \ p}{\partial r \ \partial
\theta}=\frac{-\dot{M}_0 \,
C_1(\theta)}{\sin\theta}\left\{N_3(r,\theta)
\left(\frac{-1}{r^2}+\frac{8\, a\, L}{r^5}\right)+\frac{1}{r}
\left(1-\frac{2 \, a \,L}{r^3}\right) \frac{\partial
N_3(r,\theta)}{\partial r}-\frac{30 \, a \, L}{r^6}
\tilde{V}^{(\theta)}+\frac{6 \, a \, L}{r^5}
\frac{\partial \tilde{V}^{(\theta)}}{\partial r}\right.\nonumber\\[.3cm]
&&\left.\qquad\quad+\frac{L^2}{r^4 \sin^2\theta} \left(1-\frac{2 \, a
\,L}{r^3}\right)^{-1}\frac{\cot\theta}{C_1(\theta)\tilde{V}^{(\theta)}}\left[\left(4-\frac{2
\, a \,L}{r^3}\right)\frac{1}{r}\left(1-\frac{2 \, a
\,L}{r^3}\right)^{-1}+\frac{1}{\tilde{V}^{(\theta)}}\frac{\partial
\tilde{V}^{(\theta)}}{\partial r}\right]\right\}\nonumber\\[.3cm]
&&\qquad\quad\,-\left[B_{(r)} J^{(\varphi)}+r \frac{\partial
B_{(r)}}{\partial r} J^{(\varphi)}+ r B_{(r)} \frac{\partial
J^{(\varphi)}}{\partial
r}\right]\left\{1-[V^{(\varphi)}]^2\right\}+2\, V^{(\varphi)}
\frac{\partial V^{(\varphi)}}{\partial r} r B_{(r)} J^{(\varphi)}\nonumber\\
&&\qquad\quad-\left[J^{(\varphi)} B_{(\varphi)}+r \frac{\partial J^{(\varphi)}}{\partial r} B_{(\varphi)}+r J^{(\varphi)} \frac{\partial B_{(\varphi)}}{\partial r}\right] V^{(r)} V^{(\varphi)}\nonumber\\
&&\qquad\quad-\left[\frac{\partial V^{(r)}}{\partial r} V^{(\varphi)}+V^{(r)}\frac{\partial V^{(\varphi)}}{\partial r}\right] r J^{(\varphi)} B_{(\varphi)},
\end{eqnarray}
where
\begin{eqnarray*}
N_1(r,\theta)&=&[C_1(\theta)]^2 \frac{\partial
\tilde{V}^{(\theta)}}{\partial r}
+\frac{1}{r}C_1(\theta)\frac{\partial \tilde{V}^{(\theta)}}{\partial
\theta}+\frac{1}{r} \tilde{V}^{(\theta)}\frac{d
C_1(\theta)}{d\theta}-\frac{1}{r}
\left[\left(1-\frac{2}{r}\right)\tilde{V}^{(\theta)}+\frac{\left[V^{(\varphi)}\right]^2}{\tilde{V}^{(\theta)}}\right]\\[.2cm]
&+&\left[\frac{1}{r^2}\left(1-\frac{2}{r}\right)^{-1}-\frac{6\, a \,
L}{r^4}\left(1-\frac{2 \, a \, L}{r^3}\right)^{-1}\right]
\left[\frac{1}{\tilde{V}^{(\theta)}}-\left[C_1(\theta)\right]^2\tilde{V}^{(\theta)}\right],\\[.4cm]
\frac{\partial N_1(r,\theta)}{\partial \theta}&=& 2 \frac{d
C_1(\theta)}{d \theta} \left[C_1(\theta)\frac{\partial
\tilde{V}^{(\theta)}}{\partial r}+\frac{1}{r}\frac{\partial
\tilde{V}^{(\theta)}}{\partial \theta}\right]
+[C_1(\theta)]^2\,\frac{\partial^2\tilde{V}^{(\theta)}}{\partial
\theta\, \partial r}+\frac{1}{r}
C_1(\theta)\frac{\partial^2\tilde{V}^{(\theta)}}{\partial
\theta^2}+\frac{1}{r}\tilde{V}^{(\theta)}\frac{d^2
C_1(\theta)}{d\theta^2}\\[.2cm]
&-&\frac{1}{r} \left(1-\frac{2}{r}\right)\frac{\partial
\tilde{V}^{(\theta)}}{\partial\theta}-\frac{2}{r}\frac{V^{(\varphi)}}{\tilde{V}^{(\theta)}}\frac{\partial
V^{(\varphi)}}{\partial\theta}+\frac{1}{r}\left[\frac{V^{(\varphi)}}{\tilde{V}^{(\theta)}}\right]^2\frac{\partial
\tilde{V}^{(\theta)}}{\partial\theta}\\[.2cm]
&-&\left[\frac{1}{r^2}\left(1-\frac{2}{r}\right)^{-1}-\frac{6\, a \,
L}{r^4}\left(1-\frac{2 \, a
\,L}{r^3}\right)^{-1}\right]\left[\left\{\frac{1}{[\tilde{V}^{(\theta)}]^2}+[C_1(\theta)]^2\right\}
\frac{\partial \tilde{V}^{(\theta)}}{\partial\theta}-2\, C_1(\theta)
\tilde{V}^{(\theta)} \frac{d C_1(\theta)}{d\theta}\right],
\end{eqnarray*}
\begin{eqnarray*}
N_2(r,\theta)&=&\left(1-\frac{2}{r}\right)^{-1/2}\left\{\left[\frac{\partial
B_{(\theta)}}{\partial \theta} J^{(\varphi)}+ B_{(\theta)}
\frac{\partial J^{(\varphi)}}{\partial
\theta}\right]\left\{1-[V^{(\varphi)}]^2\right\}-2
V^{(\varphi)}\frac{\partial
V^{(\varphi)}}{\partial\theta}B_{(\theta)} J^{(\varphi)}\right\}\\[.2cm]
&+&\left[\frac{\partial J^{(\varphi)}}{\partial\theta} B_{(\varphi)}+J^{(\varphi)}\frac{\partial B_{(\varphi)}}{\partial \theta}\right] \tilde{V}^{(\theta)} V^{(\varphi)}+\left[\frac{\partial V^{(\theta)}}{\partial \theta} V^{(\varphi)}+\tilde{V}^{(\theta)} \frac{\partial V^{(\varphi)}}{\partial \theta} \right]J^{(\varphi)}B_{(\varphi)},\\[.4cm]
N_3(r,\theta)&=&\frac{\partial \tilde{V}^{(\theta)}}{\partial
r}+\frac{1}{r^2}\left(1-\frac{2}{r}\right)^{-1/2}
\tilde{V}^{(\theta)}+\frac{1}{r C_1(\theta)}\frac{\partial
\tilde{V}^{(\theta)}}{\partial
\theta}+\left(1-\frac{3}{r}\right)\left(1-\frac{2}{r}\right)^{-1}\frac{\tilde{V}^{(\theta)}}{r},\\[.4cm]
\frac{\partial N_3(r,\theta)}{\partial r}&=&\frac{\partial^2
\tilde{V}^{(\theta)}}{\partial r^2}
+\frac{\tilde{V}^{(\theta)}}{r^2}\left(1-\frac{2}{r}\right)^{-2}\left[-1+\frac{6}{r}\left(1-\frac{1}{r}\right)
+\frac{1}{r}\left(1-\frac{2}{r}\right)^{1/2}\left(-2+\frac{3}{r}\right)\right]\\[.2cm]
&+&\frac{1}{r C_1(\theta)}\left(-\frac{1}{r}\frac{\partial
\tilde{V}^{(\theta)}}{\partial \theta}+\frac{\partial^2
\tilde{V}^{(\theta)}}{\partial r \partial
\theta}\right)+\frac{1}{r}\left(1-\frac{2}{r}\right)^{-1}\frac{\partial
\tilde{V}^{(\theta)}}{\partial
r}\left[\frac{1}{r}\left(1-\frac{2}{r}\right)^{1/2}+\left(1-\frac{3}{r}\right)\right].
\end{eqnarray*}
Integrability condition for pressure
\begin{eqnarray}\label{pressure_integ_cond}
\frac{\partial^2 \ p}{\partial \theta \ \partial r}=\frac{\partial^2
\ p}{\partial r  \ \partial \theta},
\end{eqnarray}
gives a second-order ordinary differential equation for
$C_1(\theta)$. In order to ready this differential equation for
computer code to solve it numerically, we arrange it in the form
$$\frac{d^2 C_1(\theta)}{d\theta^2}=\Psi(r,\theta).$$
It means that we must sort the differential equation
(\ref{pressure_integ_cond}) in terms of $\frac{d^2
C_1(\theta)}{d\theta^2}$. This term, not only appears in the fifth
term in $\frac{\partial N_1(r,\theta)}{\partial \theta}$ clearly,
but also $\frac{\partial^2 \tilde{V}^{(\theta)}}{\partial \theta^2}$
includes it. Thus, equation (\ref{pressure_integ_cond}) may be
rewritten as
\begin{eqnarray*}
\frac{-\dot{M}_0}{\left(1-\frac{2}{r}\right) \left(1-\frac{2 \, a \,
L}{r^3}\right)^{-1} r^2
\sin\theta}\left\{N_4(r,\theta)+\frac{1}{r}\left[\psi (r,\theta)\
C_1(\theta)+\tilde{V}^{(\theta)}\right]\frac{d^2
C_1(\theta)}{d\theta^2}\right\}+N_2(r,\theta)=\frac{\partial^2
\,p}{\partial r \
\partial \theta},
\end{eqnarray*}
wherein $N_4(r,\theta)=-\cot\theta \ N_1(r,\theta)+$ all the terms
of $\frac{\partial N_1(r,\theta)}{\partial \theta}$ exclude the
terms including $\frac{d^2 C_1(\theta)}{d\theta^2}$. Thus,
\begin{eqnarray*}
\Psi(r,\theta)=\frac{-\frac{r^2 \sin\theta}{\dot{M}_0}
\left(1-\frac{2}{r}\right)\left(1-\frac{2 \, a
\,L}{r^3}\right)^{-1}\left[\frac{\partial^2 \,p}{\partial r
\partial \theta}-N_2(r,\theta)\right]-N_4(r,\theta)}{\frac{1}{r}\left[\psi(r,\theta)\
C_1(\theta)+\tilde{V}^{(\theta)}\right]},
\end{eqnarray*}
here
\begin{eqnarray*}
\psi(r,\theta)=-\frac{S_0}{2}\frac{Y \
I^{3/2}}{1-[V^{(\varphi)}]^2}\left[2\, S_1 A_1 \, r
\left(1-\frac{2}{r}\right)^{-1/2}+2 \left(S_0 Y\right)^2
C_1(\theta)\right].
\end{eqnarray*}
\end{appendix}

\twocolumn


\end{document}